\documentclass[aps,prl,twocolumn]{revtex4-1}
\usepackage{amsmath}
\usepackage{amssymb}    %
\usepackage{amsthm}    %
\usepackage{epsfig}
\usepackage{subfigure}
\usepackage{float}
\usepackage{verbatim} 
\usepackage{pstricks}
\usepackage{color}
\usepackage{slashed}
\usepackage{multirow}
\usepackage{pstricks}
\usepackage{booktabs}
\usepackage{subfigure}
\usepackage{epstopdf}
\usepackage{braket}

\renewcommand{\sec}[1]{{\textbf{\textit{#1}}}.---}

\allowdisplaybreaks[3]

\def \bal#1\eal  {\begin{align} #1 \end{align}}
\newcommand{\be} {\begin{equation}}
\newcommand{\ee} {\end{equation}}

\newcommand{\ud} {\mathrm{d}}

\newcommand{\mc} {\mathcal}
\newcommand{\mb} {\mathbb}

\begin{document}


\title{Convex Geometry Perspective to the (Standard Model)
Effective Field Theory Space}
\author{Cen Zhang}
\email{cenzhang@ihep.ac.cn}
\affiliation{
Institute for High Energy Physics, and School of Physical Sciences, University
of Chinese Academy of Sciences, Beijing 100049, China
}
\affiliation{Center for High Energy Physics, Peking University, Beijing 100871, China}
\author{Shuang-Yong Zhou}
\email{zhoushy@ustc.edu.cn}
\affiliation{
Interdisciplinary Center for Theoretical Study, University of Science and Technology of China, Hefei, Anhui 230026, China}
\affiliation{Peng Huanwu Center for Fundamental Theory, Hefei, Anhui 230026, China}

\begin{abstract} We present a convex geometry perspective to the Effective
	Field Theory (EFT) parameter space.  We show that the second $s$
	derivatives of the forward EFT amplitudes form a convex cone, whose
	extremal rays are closely connected with states in the UV theory.  For
	tree-level UV completions, these rays are simply theories with all UV
	particles living in at most one irreducible representation of the
	symmetries of the theory.  In addition, all the extremal rays are
	determined by the symmetries and can be systematically identified via
	group theoretical considerations.  The implications are twofold.
	First, geometric information encoded in the EFT space can help
	reconstruct the UV completion.  In particular, we will show that the
	dim-8 operators are important in reverse engineering the UV physics
	from the Standard Model EFT, and thus deserve more theoretical and
	experimental investigations.  Second, theoretical bounds on the Wilson
	coefficients can be obtained by identifying the boundaries of the cone
	and are, in general, stronger than the current positivity bounds. We show
	explicit examples of these new bounds and demonstrate that they
	originate from the scattering amplitudes corresponding to entangled
	states.  \end{abstract} 

\maketitle

\sec{Introduction} Effective field theory (EFT) is an important framework to
systematically parameterize new high-scale phenomena.  Absent any clear
signature of new particles from the LHC data, the Standard Model EFT (SMEFT)
\cite{Weinberg:1978kz, Buchmuller:1985jz,Leung:1984ni} has become a standard
tool for studying indirect signs of new physics.  If EFT operators are detected
and the corresponding Wilson coefficients measured, the next step is to pin
down the underlying UV theory. While determining the Wilson coefficients from a
given UV theory is a systematized procedure
\cite{Henning:2014wua,Drozd:2015rsp,Henning:2016lyp,
Ellis:2016enq,Fuentes-Martin:2016uol,Zhang:2016pja,Ellis:2017jns,Kramer:2019fwz,deBlas:2017xtg,Criado:2017khh,Bakshi:2018ics},
this inverse problem can be highly nontrivial, as one set of coefficients can
be UV-completed in many ways.

A geometric perspective provides hints to this problem.  Consider the subspace
of the EFT parameters\;\footnote{By an EFT we mean a set of operators with
certain Wilson coefficients. The parameter space of (or simply space of) the
EFT is spanned by possible values of all coefficients.}, spanned by the
operators that contribute to the second $s$ derivatives of the forward 2-to-2
scattering amplitude.  The Wilson coefficients are subject to positivity bounds
\cite{Adams:2006sv} (see \cite{Pham:1985cr, Pennington:1994kc, Ananthanarayan:1994hf, Comellas:1995hq, Manohar:2008tc, Bellazzini:2016xrt, deRham:2017avq, deRham:2017zjm, nimahuanghuang} for earlier works and recent generalizations; also see the applications in SMEFT \cite{Zhang:2018shp, Bi:2019phv, Bellazzini:2018paj, Remmen:2019cyz, Remmen:2020vts} and other areas \cite{deRham:2018qqo, deRham:2017imi, Baumann:2015nta, Bellazzini:2015cra, Cheung:2016yqr,  Cheung:2016wjt, Bellazzini:2017fep, Bonifacio:2016wcb, Hinterbichler:2017qyt, Bonifacio:2017nnt, Bellazzini:2017bkb,  Bonifacio:2018vzv, Bellazzini:2019xts, Melville:2019wyy, deRham:2019ctd, Alberte:2019xfh, Alberte:2019zhd, Ye:2019oxx, Herrero-Valea:2019hde, Wang:2020jxr})
for the EFT to have a UV completion that satisfies the axiomatic principles of
quantum field theory.  These bounds on dim-8 operators are a set of linear
homogeneous inequalities of the coefficients.  The solutions form a {\it convex
cone} whose vertex is the origin of the (linear) space spanned by the
coefficients.  In this Letter, we establish a connection between the geometry
of the $s^2$-subspace of EFT and the UV physics behind. On the geometry side,
the physical space is a convex cone that can be generated as positively
weighted sums of its edges, i.e.~its {\it extremal rays} (ERs).  On the physics
side, an ER corresponds to an irreducible representation (irrep) under the
symmetries of the theory, and can be obtained {\it only} by integrating out
heavy states from this single irrep.  This geometric view helps determine the
UV physics from measurements. By using the convex nature of the subspace, one
can often draw striking conclusions about the existence of states including
their quantum numbers and couplings.  

In SMEFT, dim-8 operators
\cite{Henning:2015alf,Murphy:2020rsh,Li:2020gnx,Remmen:2019cyz} linearly
furnish this subspace.  While dim-6 coefficients are expected to be more
accurately measured, they alone are insufficient to determine UV
models: There is an infinite number of models, or combinations of UV states,
that leave no net dim-6 effect.  A UV model can be determined only modulo the
addition of these combinations. This is in contrast to dim-8, as positivity
bounds imply that all UV completions must have dim-8
effects \cite{Adams:2006sv, Zhang:2018shp}.  The dim-8 operators have attracted
increasing attention as the LHC has accumulated more and more data. Various
motivations for going beyond dim-6 have been discussed, e.g.~in 
Refs.~\cite{Liu:2016idz,Azatov:2016sqh,Bellazzini:2017bkb,Ellis:2018cos,Bellazzini:2018paj,Hays:2018zze,Ellis:2019zex,Remmen:2020vts,Alioli:2020kez}.
A number of dim-8 coefficients can be tested at the TeV level at the LHC
\cite{Ellis:2018cos,Alioli:2020kez,Sirunyan:2019der,CMS:2020meo,Sirunyan:2020tlu},
while better sensitivities are expected at future colliders
\cite{Azzi:2019yne,Ellis:2019zex}.
Furthermore, observables and opportunities that allow disentangling
dim-8 effects from the dim-6 ones exist and are being studied
\cite{Murphy:2020rsh,Alioli:2020kez,Gupta}.
We will show that the geometric connection to the UV physics gives another
important motivation to study dim-8 operators: Their coefficients contain
vital information for a bottom-up reconstruction of UV physics. 

To formulate this mapping between ERs and UV states, an accurate description of
the EFT cone is mandatory.  The current positivity bound approach is not
sufficient.  Instead, we will take a different approach that follows the
extremal representation \cite{KM} of convex cones.  Before proceeding, it is
instructive to introduce some basic concepts and facts in convex geometry.

A convex cone is a subset of a linear space that is closed under additions and
positive scalar multiplications. An extremal ray (ER) of a convex cone $\mc
C_0$ is an element $x \in\mc C_0$ that cannot be split into two other elements
in a nontrivial way, i.e.~if we write $x=y_1+y_2$ with $y_1,y_2\in \mc C_0$, we
must have $x=\lambda y_1$ or $x=\lambda y_2$, $\lambda$ being real constant.
For example, the ERs of a polyhedral cone are its edges. The dual cone $\mc
C_0^*$ of $\mc C_0$ is the set $\mc C_0^*\equiv\left\{y\, |\,  x\cdot y\ge0,
\forall x \in \mc C_0\right\}$, where $\cdot$ means the inner product of two
vectors. We have $({\mc C}_0^*)^*={\mc C}_0$, and $\mc C_1\subset \mc C_2$
implies $\mc C_1^*\supset \mc C_2^*$.  The full set of positive linear
combinations of elements in some set $\mc X$ form a convex cone, denoted by
cone($\mc X$). Its ERs are a subset of $\mc X$.

\sec{EFT amplitudes as convex cones}  
Consider the forward scattering amplitude $M_{ij\to kl}(s,t=0)$, where $s,t$
are the standard Mandelstam variables and $1\le i,j,k,l\le n$ represent the
low-energy modes.  Using analyticity of $M_{ij\to kl}(s)$ and the generalized
optical theorem, we have the following dispersion relation
\begin{flalign}
\label{eq:1first}
	{M}^{ijkl}\!=\!&\int_{(\epsilon\Lambda)^2}^{\infty}\! 
	\frac{\ud \mu\,{\rm Disc} M_{ij\to kl}(\mu)}{2i\pi(\mu-\frac{M^2}{2})^3}
	+(j\!\leftrightarrow\! l) + c.c.  
 \\
=&  \int_{(\epsilon\Lambda)^2}^{\infty}  \!\sum_{X}\!{}' \!\sum_{K=R,I}  \!\frac{\ud \mu \, {m_K}^{ij}_X {m_K}^{kl}_X }{\pi (\mu-\frac{M^2}{2})^3} 	  
 +(j\leftrightarrow l)  .
  \label{eq:1}
\end{flalign}
Here we have focused on particles with equal masses, $M^2$ being the
total mass squared, and the l.h.s.~is the second-order $s$ derivative of
$M_{ij\to kl}(s)$, with the low-energy discontinuity subtracted up to $\epsilon
\Lambda$, a scale smaller than the EFT cutoff (see Appendix for
more details and cases with different masses).  $(j\!\leftrightarrow\! l)$
means all the previous terms with the swap $j\!\leftrightarrow\! l$.
$\sum'_{X}$ denotes the sum over possible $X$ states along with their phase
spaces, and we have written the $ij\to X$ amplitude $M_{ij\to X} \equiv
{m_R}^{ij}_X + i\, {m_I}^{ij}_X$. 

The elastic version of this relation ($i=k,j=l$) has been widely used to derive
positivity bounds (because ${m_K}^{ij}_X{m_K}^{ij}_X\ge0$; see, e.g.,
\cite{Adams:2006sv}). One may also mix different polarizations
\cite{Cheung:2016yqr,deRham:2018qqo, Zhang:2018shp,Bi:2019phv} and different
particles
(e.g.~\cite{Bellazzini:2018paj,Zhang:2018shp,Bi:2019phv,Remmen:2019cyz,Remmen:2020vts,
Wang:2020jxr, Andriolo:2020lul}), to get more bounds by using
$M^{ijkl}u^iv^ju^kv^l\ge0$ (because $u^iv^j u^k
v^l{m_K}^{ij}_X{m_K}^{kl}_X=(u^i m^{ij}_{K_X} v^j)^2\ge0$), where $u^i$ and
$v^j$ enumerate the particles and polarizations \cite{longpaper}.  This can be
viewed as the positivity bound from superposed states $u^i\ket{i}$ and
$v^j\ket{j}$.  In any case, the $M^{ijkl}$ on the l.h.s.~is a low-energy
quantity and can be expressed in terms of the Wilson coefficients, either at
tree level or loop level, and we will use it as a proxy of the EFT space.  At
the tree level, $M^{ijkl}$ can be linearly mapped to the dim-8 coefficient
space \cite{Zhang:2018shp, Bi:2019phv, Bellazzini:2018paj, Remmen:2019cyz,
Remmen:2020vts}, so in the SMEFT discussions we will not distinguish the two.
Note that since our discussion will be based on $M^{ijkl}$ which is a physical
object, field redefinitions and renormalization group (RG) running will not
change our conclusions.  The approach is generically applicable to any EFT,
including the Higgs EFT, in case the latter is needed to describe $M^{ijkl}$.

Our goal is a more accurate characterization of the set $\mc C$ of all possible
$M^{ijkl}$. The main observation is that Eq.~(\ref{eq:1}) defines $\mc C$ as a
convex cone.  To see this, note that Eq.~(\ref{eq:1}) represents a positively
weighted sum of ${m_{K}}_X^{ij}{m_{K}}_X^{kl}+(j \leftrightarrow l)$, with
integration regarded as a limit of summation.  For a model-independent EFT,
${m_{K}}_X^{ij}$ are arbitrary ${n}\times n$ real matrices.  Thus the set $\mc
C$ can be viewed as a convex cone
\begin{flalign}
	\mc C=\mathrm{cone}\left(\left\{M\ \middle| \ M^{ijkl}=m^{i(j}m^{|k|l)},
	m\in \mb R^{n^2} 
\right\}\right) ,
\label{eq:cone}
\end{flalign}
i.e.~$\mc C$ is positively generated from all tensors of the form
$m^{i(j}m^{|k|l)}$, where $i(j|k|l)$ means $j,l$ indices are symmetrized.
Furthermore, $\mc C$ is a {\it salient cone}, i.e.~if $c \in \mc C,\ c\neq0,$
then $-c\notin \mc C$.  This is because any nonzero element of $\mc C$, after
contracting with $\delta^{ik}\delta^{jl}$, is positive as $m^{ij}m^{ij}>0$.
According to the Krein-Milman theorem \cite{KM}, $\mc C$ is then determined by the
convex hull of its ERs, which leads to the extremal representation of $\mc C$. 

Before moving forward, we comment on the incompleteness of the elastic
positivity bounds from superposed
states.  As they
are derived using
$M^{ijkl}u^iv^ju^kv^l\ge0$, these bounds describe the dual cone of $\mc Q\equiv
\mathrm{cone}(\{u^iv^ju^kv^l\})$. If $\mc Q=\mc C^*$, then $\mc Q^*$ is an
accurate description of $\mc C$. However, we will show explicit examples
where $\mc C^*$ contains more elements than $\mc Q$, which implies that elastic
bounds are not tight.  In this respect, finding the extremal representation of
$\mc C$ is a better approach.

\sec{ERs and UV states} 
The ERs can be found by using symmetries. The forward scattering is invariant
under an $SO(2)$ rotation around the forward direction.  Taking the SM as an
example, we can rewrite the r.h.s.~of Eq.~(\ref{eq:1}), choosing the
intermediate states $X$ as irreps (denoted by $\mathbf r$) under the $SO(2)$
rotation and the $SU(3)_C\times SU(2)_L\times U(1)_Y$ symmetries.  The
Wigner-Eckart theorem dictates that $M(ij\to X^\alpha)$ can be written as
$\braket{X|\mathcal{M}|\mathbf{r}}C_{i,j}^{r,\alpha}$, where $\alpha$ labels
the states of $\mathbf r$ and $C_{i,j}^{r,\alpha}$ is the Clebsch-Gordan (CG)
coefficients for the direct sum decomposition of
$\mathbf{r}_i\otimes\mathbf{r}_j$, with $\mathbf{r}_i(\mathbf{r}_j)$ the irrep
of $i(j)$.  The dynamics is contained in
$\braket{X|\mathcal{M}|\mathbf{r}}$, independent of $\alpha$. Equation~(\ref{eq:1})
becomes:
\begin{flalign}
	M^{ijkl}= \int_{(\epsilon\Lambda)^2}^{\infty}\ud\mu
	{\sum_{X\text{ in }\mathbf{r}}}'
	\frac{|\braket{X|\mathcal{M}|\mathbf{r}}
	|^2 }{\pi\left( \mu-\frac{1}{2}M^2 \right)^3}
	P_r^{i(j|k|l)}
	\label{eq:2}
\end{flalign}
where $P_r^{ijkl}\equiv \sum_\alpha C^{r,\alpha}_{i,j}(C^{r,\alpha}_{k,l})^*$
are the projective operators of the ${\bf r}$ representation. Similar to
Eq.~(\ref{eq:cone}), we identify the cone $\mc C$ as $
\mathrm{cone}\big(\big\{P_r^{i(j|k|l)}\big\}\big)$,
and its ERs are a subset of
$\big\{P_r^{i(j|k|l)}\big\}$.  These $j,l$-symmetrized projectors are not
necessarily extremal, so we call them potential ERs (PERs); taking their convex
hull identifies the true ERs among them. $\mc C$ is determined by the ERs.

The ERs are closely related to UV completions.  For a physics amplitude
$M^{ijkl}$ to be extremal, on the r.h.s.~of Eq.~(\ref{eq:2}), {\it only one
irrep can exist}; otherwise, $M^{ijkl}$ can be written as a sum of two different
elements of $\mc C$, which is nonextremal.  This contains important
information about the UV dynamics. For tree-level UV completions, an ER implies
that its entire $M^{ijkl}$ can be generated from the exchange of a single
(multiplet) particle, i.e.~the theory is a ``one-particle extension'' of the
SM.  It may be generated by several particles, but they must all live in the
same irrep, and have the same interaction.  For loop-level UV completions,
similarly, all multi-particle intermediate states (which may include SM
particles if RG effects are not negligible) have to live in a single irrep.
For nonperturbative UV completions, subtleties may arise, but
a similar inference might exist, if $M^{ijkl}$ can be interpreted as
coming from the exchange of UV states. We, however, leave the nonperturbative
cases for a future discussion.
More generally, any point in $\mc C$ is a positive sum of the ERs, and this
coincides with the decomposition of the intermediate UV states into irreps.
Therefore geometric information in $\mc C$ helps UV reconstruction.

This approach can be applied to subsets of particles closed under
all symmetries. The PERs continue to be projective in this subspace, so
results derived (such as bounds) are valid in general.  In the
following we will illustrate our approach with three subsets of SM fields:
scalars, vectors and fermions.  For SM particles living in one
multiplet, the number of PERs is finite, and $\mc C$ is polyhedral following a
theorem by Minkowski and Weyl \cite{Minkowski,Weyl}, which are easy to
obtain.  If more particles are involved, one may resort to more efficient
numerical algorithms, such as the reverse search algorithm \cite{Avis,lrs} for
obtaining bounds, or simply classical linear programming methods, for
testing the inclusion of given points \cite{longpaper}.

\sec{The Higgs triangular cone}
The SM Higgs boson lives in the $\mathbf 2$ of $SU(2)_L$ and carries
hypercharge $1/2$. To find the PERs, we work with real scalars, define
\begin{flalign}
	H=\left(\begin{array}{cc}
		\phi_2+i\phi_1\\
		\phi_4-i\phi_3
		\end{array}
	\right), ~
	C=\left(\begin{array}{cc}
		0 & \mathbf{1}_{2\times2} \\
		-\mathbf{1}_{2\times2} & 0 \\
	\end{array} \right) ,
\end{flalign}
and use the $\gamma$ matrices defined in Ref.~\cite{Helset:2018fgq}. The
projectors of the irreps from $\mathbf 2\otimes \mathbf 2$ define the following
PERs:
\begin{flalign}
&E_{1}^{ijkl}=\frac{1}{2}\left[C^{i(j}C^{|k|l)}+(C\gamma_4)^{i(j}(C\gamma_4)^{|k|l)}\right] ,
\nonumber\\
&E_{1S}^{ijkl}=\mathbf{1}_{4\times4}^{i(j} \mathbf{1}_{4\times4}^{|k|l)},\ E_{1A}=\gamma_4^{i(j} \gamma_4^{|k|l)} ,
\nonumber\\
&E_{3}^{ijkl}=
\frac{1}{2}\left[(C\gamma_I)^{i(j}(C\gamma_I)^{|k|l)}+(C\gamma_4\gamma_I)^{i(j}(C\gamma_4\gamma_I)^{|k|l)}\right]
\nonumber\\
& E_{3S}^{ijkl}=(\gamma_4\gamma_I)^{i(j} (\gamma_4\gamma_I)^{|k|l)},\
E_{3A}^{ijkl}=(\gamma_I)^{i(j} (\gamma_I)^{|k|l)} ,
\label{eq:rays1}
\end{flalign}
where the subscripts ${}_{1,3}$ denote the $\bf 1$ and $\bf 3$, respectively, and
${}_{S,A}$ denote the exchange symmetry of the irrep. $I$ runs from 1 to 3.
$E_1$ and $E_3$ consist of two terms, as required by hypercharge
conservation.  The UV particle for each irrep can be easily identified,
e.g.~as in Ref.~\cite{Low:2009di}.
\begin{figure}[h]
	\begin{center}
		\includegraphics[width=\linewidth]{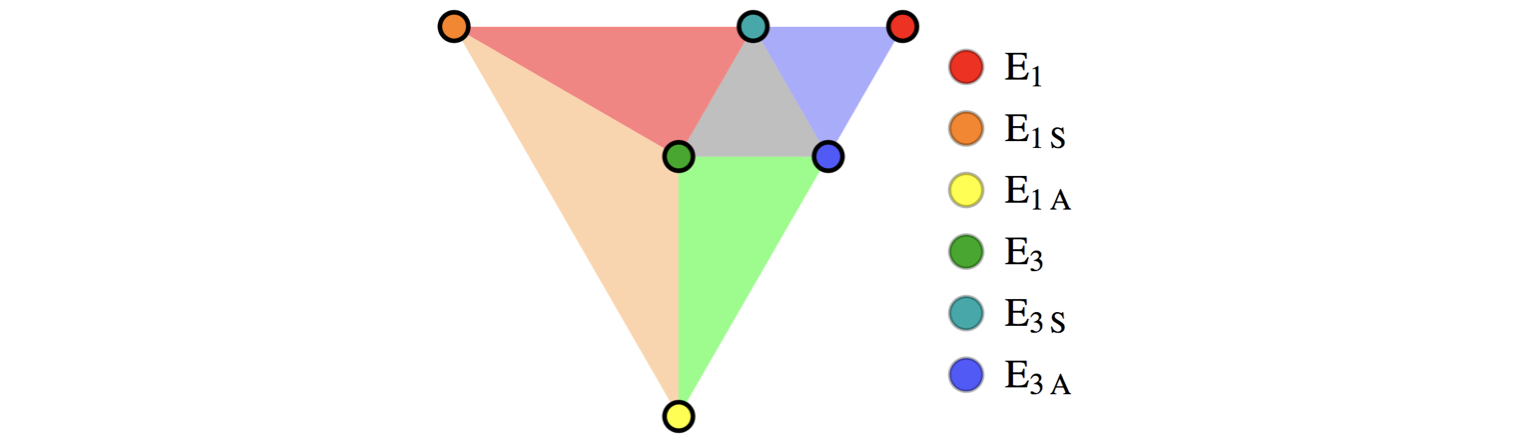}
	\end{center}
	\caption{A cross section of the Higgs triangular cone with the PERs, taken to be
		perpendicular to the direction $E_{1}+E_{1S}+E_{1A}$.}
	\label{fig:HH}
\end{figure}

Only 3 of the 6 PERs are linearly independent, as there are only 3 independent
$H^4D^4$-type operators, conventionally taken to be $O_{S,n}$, $n=0,1,2$,
defined in \cite{Degrande:2013rea}.  The convex hull of the PERs determines
$\mc C$ as a 3D triangular cone, whose cross section is shown in
Figure~\ref{fig:HH}.  There are 3 ERs: $E_{1}$, $E_{1S}$ and $E_{1A}$.  What
can we learn from this cone? First, any UV-completable EFT must stay within
this cone. Its 3 facets are, after matching to the Wilson coefficients:
$C_{S,0}\ge0$, $C_{S,0}+C_{S,2}\ge0$ and $C_{S,0}+C_{S,1}+C_{S,2}\ge0$,
$C_{S,n}$ being the coefficients of $O_{S,n}$.  These are precisely the
positivity bounds obtained from elastic scatterings of superposed Higgs modes,
albeit numerically \cite{Remmen:2019cyz}. Here we see that they are the
strongest bounds, even going beyond elastic scatterings. (This, however, is not
always true; see the $W$-boson case.) Second, the shape of the cone
contains nontrivial information about the UV completion.  Suppose the
coefficients are experimentally measured and fall into the blue region.  We can
immediately deduce that a new particle (or a multi-particle state, for
loop-level UV completions), which is a $SU(2)_L$ singlet and has hypercharge 1,
must exist and couple to $HH$, in order to generate $E_1$, because the convex
hull of all other PERs does not contain this point.  Similarly, if it falls in
the red (green) or orange region, we know that a new particle that lives in the
$1S$ ($1A$) representation must exist.

\sec{The $W$-boson polyhedral cone}
Our second example is the $W$-boson, which has 2 polarization modes and is charged
under the $\mathbf{3}$ of $SU(2)_L$. The projection operators
for $\mathbf{3}\otimes\mathbf{3}=\mathbf{1}\oplus\mathbf{3}\oplus\mathbf{5}$ of $SU(2)_L$ are:
\begin{flalign}
	&P^1_{\alpha\beta\gamma\sigma}=\frac{1}{N}\delta_{\alpha\beta}\delta_{\gamma\sigma},
	\ 
	P^2_{\alpha\beta\gamma\sigma}=\frac{1}{2}\left( \delta_{\alpha\gamma}\delta_{\beta\sigma}
	-\delta_{\alpha\sigma}\delta_{\beta\gamma}\right) ,
	\nonumber\\
	&P^3_{\alpha\beta\gamma\sigma}=\frac{1}{2}\left(
	\delta_{\alpha\gamma}\delta_{\beta\sigma}
	+\delta_{\alpha\sigma}\delta_{\beta\gamma}\right)
	-\frac{1}{N}\delta_{\alpha\beta}\delta_{\gamma\sigma} ,
\end{flalign}
where $N=3$.  For the $SO(2)$ rotation around the forward direction, the
projectors for
$\mathbf{2}\otimes\mathbf{2}=\mathbf{1}\oplus\mathbf{1}\oplus\mathbf{2}$ are
similar but with $N=2$.  With these we can construct 9 PERs, denoted as
$E_{m,n}$, from the tensor product of the $m$-th $SO(2)$ and the $n$-th
$SU(2)_L$ projectors.  5 of them are linearly independent.  All
except for $E_{3,3}$ are extremal.  This immediately determines $\mc C$
as a 5D polyhedral cone with 8 edges.

This example remarkably illustrates the efficiency of the extremal approach in
constraining the physical EFT space. To compare with the positivity bound
approach, we switch to the inequality representation and, 
after mapping to the operator coefficients, obtain:
\begin{flalign}
	&C_{T,2}\ge0,\ 
	4C_{T,1}+C_{T,2}\ge0,\
	\\
	& C_{T,2}+8C_{T,10}\ge0,\ 
	8C_{T,0}+4C_{T,1}+3C_{T,2}\ge0,\ 
	\\
	&12C_{T,0}+4C_{T,1}+5C_{T,2}+4C_{T,10}\ge0,\ 
	\label{eq:new1}
	\\
	&4C_{T,0}+4C_{T,1}+3C_{T,2}+12C_{T,10}\ge0.
	\label{eq:new2}
\end{flalign}
Again, the corresponding operators $O_{T,n}$ are defined in
Refs.~\cite{Degrande:2013rea} \footnote{$C_{T,10}$ denotes the coefficient
of $O_2^{W^4}$ of \cite{Remmen:2019cyz}, multiplied by $g_2^4/4$.}.
All these bounds except for $C_{T,2}\ge0$ have not appeared previously in the
literature, and are indeed stronger than those presented in Refs.~\cite{Bi:2019phv,
Remmen:2019cyz}. 
These coefficients parameterize the anomalous quartic-gauge-boson couplings,
currently being measured at the LHC
\cite{Sirunyan:2019der,CMS:2020meo,Sirunyan:2020tlu}, so they alone are
important results.  The first four bounds can be identified as positivity
bounds by scattering various superposed states of $\ket{W^{1,2}_{x,y}}$
[superscripts for $SU(2)_L$ and subscripts for polarization].
The last two bounds, Eqs.~(\ref{eq:new1}) and (\ref{eq:new2}), deserve
more attention: They cannot be derived from any elastic scattering between
superposed states, so they are beyond elastic positivity.

\sec{More than elastic positivity} As explained already,
elastic positivity fails to give a complete description of
$\mc C$, because, in general, $\mathcal{C}^*$ contains more elements than $\mathcal{Q}$.
The two bounds in Eqs.~(\ref{eq:new1}) and (\ref{eq:new2})
are indeed from the following elements of $\mc C^*$, not contained in $\mathcal{Q}$:
\begin{flalign}
	&T_1=6E_{1,1}+3E_{2,1}+6E_{2,2}+3/2E_{3,1}+3E_{3,3}
	\label{eq:T1}
	\\
	&T_2=5/2E_{1,1}+5E_{1,2}+E_{1,3}+15/2E_{2,1}+3E_{3,3}  .
	\label{eq:T2}
\end{flalign}
One can show that $T^{ijkl}_{1,2}M^{ijkl}\ge0$, which lead to
Eqs.~(\ref{eq:new1}) and (\ref{eq:new2}) respectively, and that
$T_{1,2}\notin \mc Q$, which implies that those bounds
cannot be derived from scattering between superposed states (See
the Appendix for a proof with more details).

The fact that $T_{1,2}\notin \mc Q$ suggests that the dispersion relation of
scattering amplitudes with entangled states can provide additional information
about the UV completion.  Positivity bounds would not capture this information
unless there is a systematic and efficient way to tackle all elements in $\mc
C^*$.  Note that the $T_{1,2}$ tensors are independent of this specific
problem, and may lead to new bounds also for other operators or EFTs, whenever
the number of states $n\ge6$.
Our extremal approach naturally captures all such cases.

\begin{figure}[ht]
	\begin{center}
		\includegraphics[width=\linewidth]{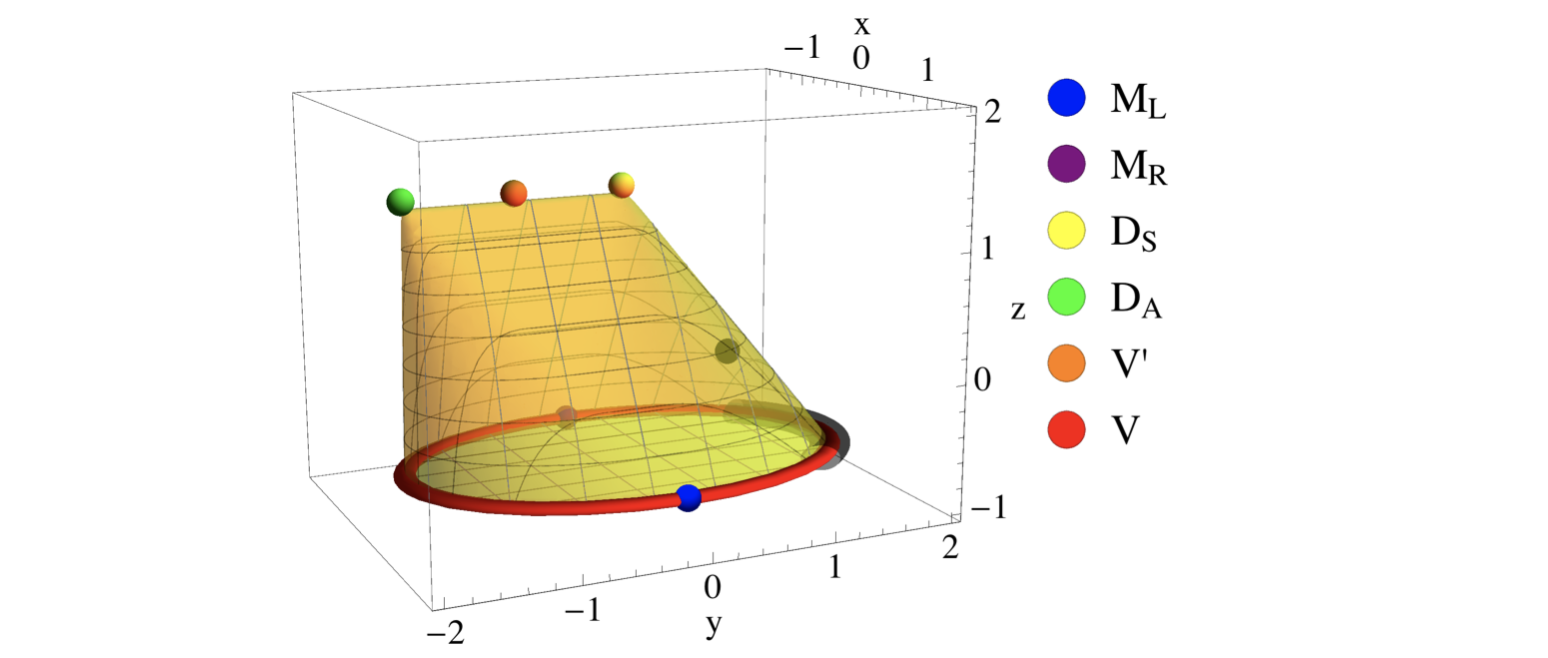}
	\end{center}
	\caption{A slice of the 4D fermion cone, taken to be perpendicular to
	the direction $(1,1,0,1)$. The three axes are taken to be
$(1,-1,0,0)$, $(0,0,1,0)$ and $(-1,-1,0,2)$.}
	\label{fig:FF}
\end{figure}

\sec{The fermion circular cone}
Lastly, we consider SM-like chiral fermions, with left- and right-handed
components carrying
different hypercharges but other symmetries neglected for simplicity.
Defining $J_{L,R}^\mu\equiv \bar f_{L,R}\gamma^\mu f_{L,R}$, we use the following
basis:
\begin{flalign}
	&O_1=-\partial^\mu J_L^\nu \partial_\mu J_{L\nu},\ 
	O_2=-\partial^\mu J_R^\nu \partial_\mu J_{R\nu},
	\nonumber\\
	&O_3=\partial^\mu J_L^\nu \partial_\mu J_{R\nu},\ 
	O_{4}=D^\mu\left( \bar f_L f_R \right) D_\mu \left( \bar f_R  f_L \right) .
\end{flalign}
We simply show the PERs, in terms of the coefficient
vector $\vec C=(C_1,C_2,C_3,C_4)$:
\begin{flalign*}
	\begin{aligned}
	&M_L:(1,0,0,0), 
	\\
	&M_R:(0,1,0,0), 
\end{aligned}
\ \ 
\begin{aligned}
	&D_S:(0,0,0,1), 
	\\
	&D_A:(0,0,-1,1), 
\end{aligned}
\ \ 
\begin{aligned}
	&V:(1,r^2,-2r,0), 
	\\
        &V':(0,0,-1,2).
\end{aligned}
\end{flalign*}
$M_{L,R}$ are from Majorana-type scalar couplings with two $f_{L}$'s or two
$f_{R}$'s. $D$ is from a Dirac-type scalar coupling, with subscripts ${}_{S,A}$
indicating the exchange symmetry.  $V$($V'$) is from the vector coupling formed
by same(opposite)-chirality fermions.  $r$ is the ratio between $R/L$
couplings. Since $V$ is continuously parameterized by $r$,
$\mc C$ has a curved boundary. 
In Figure~\ref{fig:FF} we show a 3D slice of $\mc C$.
The boundaries
are given by
$C_1,C_2,C_4\ge0$ and $2\sqrt{C_1C_2}\ge \max(C_3,-C_3-C_4)$.  

\sec{A geometric view for UV-determination} 
Let us reiterate what the Higgs example tells us in more general cases.
Let $\mc E_{\setminus a}$ be the convex hull of all PERs with one of them,
$\vec E_a$, removed.  If the measured coefficients, denoted as
$\vec{C}_\mathrm{exp}$, are not contained by $\mc
E_{\setminus a}$, then a tree-level UV completion must contain a particle
that couples with the $E_a$ irrep. This feature extends to loop-generated cases.
For example, in the blue region of Figure~\ref{fig:HH}, there must
exist some multi-particle state 
that couples to $HH$, carries hypercharge 1, and contains a $SU(2)_L$ singlet. 

Quantitative statements can be made.  For a measured
$\vec{C}_\mathrm{exp}$ in the blue region, there is
a minimum $\lambda$ such that $\vec C_\mathrm{exp}-\lambda \vec E_1\in
\mc E_{\setminus 1}$. This sets a lower bound on the strength of the UV coupling
that generates $\vec E_1$. Similarly, an upper bound can be set
using $\vec C_\mathrm{exp}-\lambda \vec E_i\in \mc C$ for all $\vec E_{i}$.
As a second example,
consider the fermion cone and assume $\vec{C}_\mathrm{exp}\propto
(1,0.8,1.4,1)$ is observed (see the black point in Figure~\ref{fig:FF}).  If
a small arc on $V$ (shown in black) 
is removed, the convex hull of remaining PERs does not contain
$\vec{C}_\mathrm{exp}$. It follows that a UV state exists and couples to the
fermions with $V/A$-type couplings, and an upper bound on the coupling ratio
$|g_V/g_A|<0.35$ can be set.  There are many other interesting and
phenomenologically relevant examples, where convex hulls can be used to
infer UV states. This is not possible at dim-6, as the PERs would positively
span the entire space.

As a final remark, we have shown that concepts and theorems in convex geometry
help develop a deeper understanding of the EFT space, to improve the
positivity bounds, and to determine the UV completion\,\footnote{The EFThedron
	of \cite{nimahuanghuang} also connects the the UV states
	in a geometric point of view. Convex objects such as
cyclic polytopes are found to constrain sequences of operators with increasing
dimensions. Here we consider EFTs endowed with symmetries and focus on
operators with lowest dimensions.}.  We hope that through this geometric
perspective, other results in convex geometry may find their applications in
particle physics.

\section{Acknowledgements}
We thank Gauthier Durieux, Jiayin Gu, Yu-tin Huang, Fabio
Maltoni, Jing Shu, Zi-Yue Wang and Ming-Lei Xiao for helpful discussions and
comments.  C.Z.~is supported by IHEP under Contract No.~Y7515540U1, and by
National Natural Science Foundation of China (NSFC) under Grant No.~12035008.
S.-Y.Z.~acknowledges support from the starting Grants from University of Science
and Technology of China under Grant No.~KY2030000089 and No.~GG2030040375, and is
also supported by NSFC under grant No.~11947301 and No.~12075233.

\bibliography{bib}

\begin{thebibliography}{76}%
\makeatletter
\providecommand \@ifxundefined [1]{%
 \@ifx{#1\undefined}
}%
\providecommand \@ifnum [1]{%
 \ifnum #1\expandafter \@firstoftwo
 \else \expandafter \@secondoftwo
 \fi
}%
\providecommand \@ifx [1]{%
 \ifx #1\expandafter \@firstoftwo
 \else \expandafter \@secondoftwo
 \fi
}%
\providecommand \natexlab [1]{#1}%
\providecommand \enquote  [1]{``#1''}%
\providecommand \bibnamefont  [1]{#1}%
\providecommand \bibfnamefont [1]{#1}%
\providecommand \citenamefont [1]{#1}%
\providecommand \href@noop [0]{\@secondoftwo}%
\providecommand \href [0]{\begingroup \@sanitize@url \@href}%
\providecommand \@href[1]{\@@startlink{#1}\@@href}%
\providecommand \@@href[1]{\endgroup#1\@@endlink}%
\providecommand \@sanitize@url [0]{\catcode `\\12\catcode `\$12\catcode
  `\&12\catcode `\#12\catcode `\^12\catcode `\_12\catcode `\%12\relax}%
\providecommand \@@startlink[1]{}%
\providecommand \@@endlink[0]{}%
\providecommand \url  [0]{\begingroup\@sanitize@url \@url }%
\providecommand \@url [1]{\endgroup\@href {#1}{\urlprefix }}%
\providecommand \urlprefix  [0]{URL }%
\providecommand \Eprint [0]{\href }%
\providecommand \doibase [0]{http://dx.doi.org/}%
\providecommand \selectlanguage [0]{\@gobble}%
\providecommand \bibinfo  [0]{\@secondoftwo}%
\providecommand \bibfield  [0]{\@secondoftwo}%
\providecommand \translation [1]{[#1]}%
\providecommand \BibitemOpen [0]{}%
\providecommand \bibitemStop [0]{}%
\providecommand \bibitemNoStop [0]{.\EOS\space}%
\providecommand \EOS [0]{\spacefactor3000\relax}%
\providecommand \BibitemShut  [1]{\csname bibitem#1\endcsname}%
\let\auto@bib@innerbib\@empty
\bibitem [{\citenamefont {Weinberg}(1979)}]{Weinberg:1978kz}%
  \BibitemOpen
  \bibfield  {author} {\bibinfo {author} {\bibfnamefont {S.}~\bibnamefont
  {Weinberg}},\ }\bibfield  {booktitle} {\emph {\bibinfo {booktitle}
  {{Proceedings, Symposium Honoring Julian Schwinger on the Occasion of his
  60th Birthday: Los Angeles, California, February 18-19, 1978}}},\ }\href
  {\doibase 10.1016/0378-4371(79)90223-1} {\bibfield  {journal} {\bibinfo
  {journal} {Physica}\ }\textbf {\bibinfo {volume} {A96}},\ \bibinfo {pages}
  {327} (\bibinfo {year} {1979})}\BibitemShut {NoStop}%
\bibitem [{\citenamefont {Buchmuller}\ and\ \citenamefont
  {Wyler}(1986)}]{Buchmuller:1985jz}%
  \BibitemOpen
  \bibfield  {author} {\bibinfo {author} {\bibfnamefont {W.}~\bibnamefont
  {Buchmuller}}\ and\ \bibinfo {author} {\bibfnamefont {D.}~\bibnamefont
  {Wyler}},\ }\href {\doibase 10.1016/0550-3213(86)90262-2} {\bibfield
  {journal} {\bibinfo  {journal} {Nucl. Phys.}\ }\textbf {\bibinfo {volume}
  {B268}},\ \bibinfo {pages} {621} (\bibinfo {year} {1986})}\BibitemShut
  {NoStop}%
\bibitem [{\citenamefont {Leung}\ \emph {et~al.}(1986)\citenamefont {Leung},
  \citenamefont {Love},\ and\ \citenamefont {Rao}}]{Leung:1984ni}%
  \BibitemOpen
  \bibfield  {author} {\bibinfo {author} {\bibfnamefont {C.~N.}\ \bibnamefont
  {Leung}}, \bibinfo {author} {\bibfnamefont {S.~T.}\ \bibnamefont {Love}}, \
  and\ \bibinfo {author} {\bibfnamefont {S.}~\bibnamefont {Rao}},\ }\href
  {\doibase 10.1007/BF01588041} {\bibfield  {journal} {\bibinfo  {journal} {Z.
  Phys.}\ }\textbf {\bibinfo {volume} {C31}},\ \bibinfo {pages} {433} (\bibinfo
  {year} {1986})}\BibitemShut {NoStop}%
\bibitem [{\citenamefont {Henning}\ \emph {et~al.}(2016)\citenamefont
  {Henning}, \citenamefont {Lu},\ and\ \citenamefont
  {Murayama}}]{Henning:2014wua}%
  \BibitemOpen
  \bibfield  {author} {\bibinfo {author} {\bibfnamefont {B.}~\bibnamefont
  {Henning}}, \bibinfo {author} {\bibfnamefont {X.}~\bibnamefont {Lu}}, \ and\
  \bibinfo {author} {\bibfnamefont {H.}~\bibnamefont {Murayama}},\ }\href
  {\doibase 10.1007/JHEP01(2016)023} {\bibfield  {journal} {\bibinfo  {journal}
  {JHEP}\ }\textbf {\bibinfo {volume} {01}},\ \bibinfo {pages} {023} (\bibinfo
  {year} {2016})},\ \Eprint {http://arxiv.org/abs/1412.1837} {arXiv:1412.1837
  [hep-ph]} \BibitemShut {NoStop}%
\bibitem [{\citenamefont {Drozd}\ \emph {et~al.}(2016)\citenamefont {Drozd},
  \citenamefont {Ellis}, \citenamefont {Quevillon},\ and\ \citenamefont
  {You}}]{Drozd:2015rsp}%
  \BibitemOpen
  \bibfield  {author} {\bibinfo {author} {\bibfnamefont {A.}~\bibnamefont
  {Drozd}}, \bibinfo {author} {\bibfnamefont {J.}~\bibnamefont {Ellis}},
  \bibinfo {author} {\bibfnamefont {J.}~\bibnamefont {Quevillon}}, \ and\
  \bibinfo {author} {\bibfnamefont {T.}~\bibnamefont {You}},\ }\href {\doibase
  10.1007/JHEP03(2016)180} {\bibfield  {journal} {\bibinfo  {journal} {JHEP}\
  }\textbf {\bibinfo {volume} {03}},\ \bibinfo {pages} {180} (\bibinfo {year}
  {2016})},\ \Eprint {http://arxiv.org/abs/1512.03003} {arXiv:1512.03003
  [hep-ph]} \BibitemShut {NoStop}%
\bibitem [{\citenamefont {Henning}\ \emph {et~al.}(2018)\citenamefont
  {Henning}, \citenamefont {Lu},\ and\ \citenamefont
  {Murayama}}]{Henning:2016lyp}%
  \BibitemOpen
  \bibfield  {author} {\bibinfo {author} {\bibfnamefont {B.}~\bibnamefont
  {Henning}}, \bibinfo {author} {\bibfnamefont {X.}~\bibnamefont {Lu}}, \ and\
  \bibinfo {author} {\bibfnamefont {H.}~\bibnamefont {Murayama}},\ }\href
  {\doibase 10.1007/JHEP01(2018)123} {\bibfield  {journal} {\bibinfo  {journal}
  {JHEP}\ }\textbf {\bibinfo {volume} {01}},\ \bibinfo {pages} {123} (\bibinfo
  {year} {2018})},\ \Eprint {http://arxiv.org/abs/1604.01019} {arXiv:1604.01019
  [hep-ph]} \BibitemShut {NoStop}%
\bibitem [{\citenamefont {Ellis}\ \emph {et~al.}(2016)\citenamefont {Ellis},
  \citenamefont {Quevillon}, \citenamefont {You},\ and\ \citenamefont
  {Zhang}}]{Ellis:2016enq}%
  \BibitemOpen
  \bibfield  {author} {\bibinfo {author} {\bibfnamefont {S.~A.~R.}\
  \bibnamefont {Ellis}}, \bibinfo {author} {\bibfnamefont {J.}~\bibnamefont
  {Quevillon}}, \bibinfo {author} {\bibfnamefont {T.}~\bibnamefont {You}}, \
  and\ \bibinfo {author} {\bibfnamefont {Z.}~\bibnamefont {Zhang}},\ }\href
  {\doibase 10.1016/j.physletb.2016.09.016} {\bibfield  {journal} {\bibinfo
  {journal} {Phys. Lett. B}\ }\textbf {\bibinfo {volume} {762}},\ \bibinfo
  {pages} {166} (\bibinfo {year} {2016})},\ \Eprint
  {http://arxiv.org/abs/1604.02445} {arXiv:1604.02445 [hep-ph]} \BibitemShut
  {NoStop}%
\bibitem [{\citenamefont {Fuentes-Martin}\ \emph {et~al.}(2016)\citenamefont
  {Fuentes-Martin}, \citenamefont {Portoles},\ and\ \citenamefont
  {Ruiz-Femenia}}]{Fuentes-Martin:2016uol}%
  \BibitemOpen
  \bibfield  {author} {\bibinfo {author} {\bibfnamefont {J.}~\bibnamefont
  {Fuentes-Martin}}, \bibinfo {author} {\bibfnamefont {J.}~\bibnamefont
  {Portoles}}, \ and\ \bibinfo {author} {\bibfnamefont {P.}~\bibnamefont
  {Ruiz-Femenia}},\ }\href {\doibase 10.1007/JHEP09(2016)156} {\bibfield
  {journal} {\bibinfo  {journal} {JHEP}\ }\textbf {\bibinfo {volume} {09}},\
  \bibinfo {pages} {156} (\bibinfo {year} {2016})},\ \Eprint
  {http://arxiv.org/abs/1607.02142} {arXiv:1607.02142 [hep-ph]} \BibitemShut
  {NoStop}%
\bibitem [{\citenamefont {Zhang}(2017)}]{Zhang:2016pja}%
  \BibitemOpen
  \bibfield  {author} {\bibinfo {author} {\bibfnamefont {Z.}~\bibnamefont
  {Zhang}},\ }\href {\doibase 10.1007/JHEP05(2017)152} {\bibfield  {journal}
  {\bibinfo  {journal} {JHEP}\ }\textbf {\bibinfo {volume} {05}},\ \bibinfo
  {pages} {152} (\bibinfo {year} {2017})},\ \Eprint
  {http://arxiv.org/abs/1610.00710} {arXiv:1610.00710 [hep-ph]} \BibitemShut
  {NoStop}%
\bibitem [{\citenamefont {Ellis}\ \emph {et~al.}(2017)\citenamefont {Ellis},
  \citenamefont {Quevillon}, \citenamefont {You},\ and\ \citenamefont
  {Zhang}}]{Ellis:2017jns}%
  \BibitemOpen
  \bibfield  {author} {\bibinfo {author} {\bibfnamefont {S.~A.~R.}\
  \bibnamefont {Ellis}}, \bibinfo {author} {\bibfnamefont {J.}~\bibnamefont
  {Quevillon}}, \bibinfo {author} {\bibfnamefont {T.}~\bibnamefont {You}}, \
  and\ \bibinfo {author} {\bibfnamefont {Z.}~\bibnamefont {Zhang}},\ }\href
  {\doibase 10.1007/JHEP08(2017)054} {\bibfield  {journal} {\bibinfo  {journal}
  {JHEP}\ }\textbf {\bibinfo {volume} {08}},\ \bibinfo {pages} {054} (\bibinfo
  {year} {2017})},\ \Eprint {http://arxiv.org/abs/1706.07765} {arXiv:1706.07765
  [hep-ph]} \BibitemShut {NoStop}%
\bibitem [{\citenamefont {Krämer}\ \emph {et~al.}(2020)\citenamefont
  {Krämer}, \citenamefont {Summ},\ and\ \citenamefont
  {Voigt}}]{Kramer:2019fwz}%
  \BibitemOpen
  \bibfield  {author} {\bibinfo {author} {\bibfnamefont {M.}~\bibnamefont
  {Krämer}}, \bibinfo {author} {\bibfnamefont {B.}~\bibnamefont {Summ}}, \
  and\ \bibinfo {author} {\bibfnamefont {A.}~\bibnamefont {Voigt}},\ }\href
  {\doibase 10.1007/JHEP01(2020)079} {\bibfield  {journal} {\bibinfo  {journal}
  {JHEP}\ }\textbf {\bibinfo {volume} {01}},\ \bibinfo {pages} {079} (\bibinfo
  {year} {2020})},\ \Eprint {http://arxiv.org/abs/1908.04798} {arXiv:1908.04798
  [hep-ph]} \BibitemShut {NoStop}%
\bibitem [{\citenamefont {de~Blas}\ \emph {et~al.}(2018)\citenamefont
  {de~Blas}, \citenamefont {Criado}, \citenamefont {Perez-Victoria},\ and\
  \citenamefont {Santiago}}]{deBlas:2017xtg}%
  \BibitemOpen
  \bibfield  {author} {\bibinfo {author} {\bibfnamefont {J.}~\bibnamefont
  {de~Blas}}, \bibinfo {author} {\bibfnamefont {J.}~\bibnamefont {Criado}},
  \bibinfo {author} {\bibfnamefont {M.}~\bibnamefont {Perez-Victoria}}, \ and\
  \bibinfo {author} {\bibfnamefont {J.}~\bibnamefont {Santiago}},\ }\href
  {\doibase 10.1007/JHEP03(2018)109} {\bibfield  {journal} {\bibinfo  {journal}
  {JHEP}\ }\textbf {\bibinfo {volume} {03}},\ \bibinfo {pages} {109} (\bibinfo
  {year} {2018})},\ \Eprint {http://arxiv.org/abs/1711.10391} {arXiv:1711.10391
  [hep-ph]} \BibitemShut {NoStop}%
\bibitem [{\citenamefont {Criado}(2018)}]{Criado:2017khh}%
  \BibitemOpen
  \bibfield  {author} {\bibinfo {author} {\bibfnamefont {J.~C.}\ \bibnamefont
  {Criado}},\ }\href {\doibase 10.1016/j.cpc.2018.02.016} {\bibfield  {journal}
  {\bibinfo  {journal} {Comput. Phys. Commun.}\ }\textbf {\bibinfo {volume}
  {227}},\ \bibinfo {pages} {42} (\bibinfo {year} {2018})},\ \Eprint
  {http://arxiv.org/abs/1710.06445} {arXiv:1710.06445 [hep-ph]} \BibitemShut
  {NoStop}%
\bibitem [{\citenamefont {Das~Bakshi}\ \emph {et~al.}(2019)\citenamefont
  {Das~Bakshi}, \citenamefont {Chakrabortty},\ and\ \citenamefont
  {Patra}}]{Bakshi:2018ics}%
  \BibitemOpen
  \bibfield  {author} {\bibinfo {author} {\bibfnamefont {S.}~\bibnamefont
  {Das~Bakshi}}, \bibinfo {author} {\bibfnamefont {J.}~\bibnamefont
  {Chakrabortty}}, \ and\ \bibinfo {author} {\bibfnamefont {S.~K.}\
  \bibnamefont {Patra}},\ }\href {\doibase 10.1140/epjc/s10052-018-6444-2}
  {\bibfield  {journal} {\bibinfo  {journal} {Eur. Phys. J. C}\ }\textbf
  {\bibinfo {volume} {79}},\ \bibinfo {pages} {21} (\bibinfo {year} {2019})},\
  \Eprint {http://arxiv.org/abs/1808.04403} {arXiv:1808.04403 [hep-ph]}
  \BibitemShut {NoStop}%
\bibitem [{Note1()}]{Note1}%
  \BibitemOpen
  \bibinfo {note} {By an EFT we mean a set of operators with certain Wilson
  coefficients. The parameter space of (or simply space of) the EFT is spanned
  by possible values of all coefficients.}\BibitemShut {Stop}%
\bibitem [{\citenamefont {Adams}\ \emph {et~al.}(2006)\citenamefont {Adams},
  \citenamefont {Arkani-Hamed}, \citenamefont {Dubovsky}, \citenamefont
  {Nicolis},\ and\ \citenamefont {Rattazzi}}]{Adams:2006sv}%
  \BibitemOpen
  \bibfield  {author} {\bibinfo {author} {\bibfnamefont {A.}~\bibnamefont
  {Adams}}, \bibinfo {author} {\bibfnamefont {N.}~\bibnamefont {Arkani-Hamed}},
  \bibinfo {author} {\bibfnamefont {S.}~\bibnamefont {Dubovsky}}, \bibinfo
  {author} {\bibfnamefont {A.}~\bibnamefont {Nicolis}}, \ and\ \bibinfo
  {author} {\bibfnamefont {R.}~\bibnamefont {Rattazzi}},\ }\href {\doibase
  10.1088/1126-6708/2006/10/014} {\bibfield  {journal} {\bibinfo  {journal}
  {JHEP}\ }\textbf {\bibinfo {volume} {10}},\ \bibinfo {pages} {014} (\bibinfo
  {year} {2006})},\ \Eprint {http://arxiv.org/abs/hep-th/0602178}
  {arXiv:hep-th/0602178 [hep-th]} \BibitemShut {NoStop}%
\bibitem [{\citenamefont {Pham}\ and\ \citenamefont
  {Truong}(1985)}]{Pham:1985cr}%
  \BibitemOpen
  \bibfield  {author} {\bibinfo {author} {\bibfnamefont {T.~N.}\ \bibnamefont
  {Pham}}\ and\ \bibinfo {author} {\bibfnamefont {T.~N.}\ \bibnamefont
  {Truong}},\ }\href {\doibase 10.1103/PhysRevD.31.3027} {\bibfield  {journal}
  {\bibinfo  {journal} {Phys. Rev.}\ }\textbf {\bibinfo {volume} {D31}},\
  \bibinfo {pages} {3027} (\bibinfo {year} {1985})}\BibitemShut {NoStop}%
\bibitem [{\citenamefont {Pennington}\ and\ \citenamefont
  {Portoles}(1995)}]{Pennington:1994kc}%
  \BibitemOpen
  \bibfield  {author} {\bibinfo {author} {\bibfnamefont {M.~R.}\ \bibnamefont
  {Pennington}}\ and\ \bibinfo {author} {\bibfnamefont {J.}~\bibnamefont
  {Portoles}},\ }\href {\doibase 10.1016/0370-2693(94)01551-M} {\bibfield
  {journal} {\bibinfo  {journal} {Phys. Lett.}\ }\textbf {\bibinfo {volume}
  {B344}},\ \bibinfo {pages} {399} (\bibinfo {year} {1995})},\ \Eprint
  {http://arxiv.org/abs/hep-ph/9409426} {arXiv:hep-ph/9409426 [hep-ph]}
  \BibitemShut {NoStop}%
\bibitem [{\citenamefont {Ananthanarayan}\ \emph {et~al.}(1995)\citenamefont
  {Ananthanarayan}, \citenamefont {Toublan},\ and\ \citenamefont
  {Wanders}}]{Ananthanarayan:1994hf}%
  \BibitemOpen
  \bibfield  {author} {\bibinfo {author} {\bibfnamefont {B.}~\bibnamefont
  {Ananthanarayan}}, \bibinfo {author} {\bibfnamefont {D.}~\bibnamefont
  {Toublan}}, \ and\ \bibinfo {author} {\bibfnamefont {G.}~\bibnamefont
  {Wanders}},\ }\href {\doibase 10.1103/PhysRevD.51.1093} {\bibfield  {journal}
  {\bibinfo  {journal} {Phys. Rev.}\ }\textbf {\bibinfo {volume} {D51}},\
  \bibinfo {pages} {1093} (\bibinfo {year} {1995})},\ \Eprint
  {http://arxiv.org/abs/hep-ph/9410302} {arXiv:hep-ph/9410302 [hep-ph]}
  \BibitemShut {NoStop}%
\bibitem [{\citenamefont {Comellas}\ \emph {et~al.}(1995)\citenamefont
  {Comellas}, \citenamefont {Latorre},\ and\ \citenamefont
  {Taron}}]{Comellas:1995hq}%
  \BibitemOpen
  \bibfield  {author} {\bibinfo {author} {\bibfnamefont {J.}~\bibnamefont
  {Comellas}}, \bibinfo {author} {\bibfnamefont {J.~I.}\ \bibnamefont
  {Latorre}}, \ and\ \bibinfo {author} {\bibfnamefont {J.}~\bibnamefont
  {Taron}},\ }\href {\doibase 10.1016/0370-2693(95)01110-C} {\bibfield
  {journal} {\bibinfo  {journal} {Phys. Lett.}\ }\textbf {\bibinfo {volume}
  {B360}},\ \bibinfo {pages} {109} (\bibinfo {year} {1995})},\ \Eprint
  {http://arxiv.org/abs/hep-ph/9507258} {arXiv:hep-ph/9507258 [hep-ph]}
  \BibitemShut {NoStop}%
\bibitem [{\citenamefont {Manohar}\ and\ \citenamefont
  {Mateu}(2008)}]{Manohar:2008tc}%
  \BibitemOpen
  \bibfield  {author} {\bibinfo {author} {\bibfnamefont {A.~V.}\ \bibnamefont
  {Manohar}}\ and\ \bibinfo {author} {\bibfnamefont {V.}~\bibnamefont
  {Mateu}},\ }\href {\doibase 10.1103/PhysRevD.77.094019} {\bibfield  {journal}
  {\bibinfo  {journal} {Phys. Rev.}\ }\textbf {\bibinfo {volume} {D77}},\
  \bibinfo {pages} {094019} (\bibinfo {year} {2008})},\ \Eprint
  {http://arxiv.org/abs/0801.3222} {arXiv:0801.3222 [hep-ph]} \BibitemShut
  {NoStop}%
\bibitem [{\citenamefont {Bellazzini}(2017)}]{Bellazzini:2016xrt}%
  \BibitemOpen
  \bibfield  {author} {\bibinfo {author} {\bibfnamefont {B.}~\bibnamefont
  {Bellazzini}},\ }\href {\doibase 10.1007/JHEP02(2017)034} {\bibfield
  {journal} {\bibinfo  {journal} {JHEP}\ }\textbf {\bibinfo {volume} {02}},\
  \bibinfo {pages} {034} (\bibinfo {year} {2017})},\ \Eprint
  {http://arxiv.org/abs/1605.06111} {arXiv:1605.06111 [hep-th]} \BibitemShut
  {NoStop}%
\bibitem [{\citenamefont {de~Rham}\ \emph
  {et~al.}(2017{\natexlab{a}})\citenamefont {de~Rham}, \citenamefont
  {Melville}, \citenamefont {Tolley},\ and\ \citenamefont
  {Zhou}}]{deRham:2017avq}%
  \BibitemOpen
  \bibfield  {author} {\bibinfo {author} {\bibfnamefont {C.}~\bibnamefont
  {de~Rham}}, \bibinfo {author} {\bibfnamefont {S.}~\bibnamefont {Melville}},
  \bibinfo {author} {\bibfnamefont {A.~J.}\ \bibnamefont {Tolley}}, \ and\
  \bibinfo {author} {\bibfnamefont {S.-Y.}\ \bibnamefont {Zhou}},\ }\href
  {\doibase 10.1103/PhysRevD.96.081702} {\bibfield  {journal} {\bibinfo
  {journal} {Phys. Rev.}\ }\textbf {\bibinfo {volume} {D96}},\ \bibinfo {pages}
  {081702} (\bibinfo {year} {2017}{\natexlab{a}})},\ \Eprint
  {http://arxiv.org/abs/1702.06134} {arXiv:1702.06134 [hep-th]} \BibitemShut
  {NoStop}%
\bibitem [{\citenamefont {de~Rham}\ \emph {et~al.}(2018)\citenamefont
  {de~Rham}, \citenamefont {Melville}, \citenamefont {Tolley},\ and\
  \citenamefont {Zhou}}]{deRham:2017zjm}%
  \BibitemOpen
  \bibfield  {author} {\bibinfo {author} {\bibfnamefont {C.}~\bibnamefont
  {de~Rham}}, \bibinfo {author} {\bibfnamefont {S.}~\bibnamefont {Melville}},
  \bibinfo {author} {\bibfnamefont {A.~J.}\ \bibnamefont {Tolley}}, \ and\
  \bibinfo {author} {\bibfnamefont {S.-Y.}\ \bibnamefont {Zhou}},\ }\href
  {\doibase 10.1007/JHEP03(2018)011} {\bibfield  {journal} {\bibinfo  {journal}
  {JHEP}\ }\textbf {\bibinfo {volume} {03}},\ \bibinfo {pages} {011} (\bibinfo
  {year} {2018})},\ \Eprint {http://arxiv.org/abs/1706.02712} {arXiv:1706.02712
  [hep-th]} \BibitemShut {NoStop}%
\bibitem [{\citenamefont {Arkani-Hamed}\ \emph {et~al.}(pear)\citenamefont
  {Arkani-Hamed}, \citenamefont {Huang},\ and\ \citenamefont
  {Huang}}]{nimahuanghuang}%
  \BibitemOpen
  \bibfield  {author} {\bibinfo {author} {\bibfnamefont {N.}~\bibnamefont
  {Arkani-Hamed}}, \bibinfo {author} {\bibfnamefont {Y.}~\bibnamefont {Huang}},
  \ and\ \bibinfo {author} {\bibfnamefont {T.-C.}\ \bibnamefont {Huang}},\
  }\href@noop {} {\  (\bibinfo {year} {{\it To appear}})}\BibitemShut {NoStop}%
\bibitem [{\citenamefont {Zhang}\ and\ \citenamefont
  {Zhou}(2019)}]{Zhang:2018shp}%
  \BibitemOpen
  \bibfield  {author} {\bibinfo {author} {\bibfnamefont {C.}~\bibnamefont
  {Zhang}}\ and\ \bibinfo {author} {\bibfnamefont {S.-Y.}\ \bibnamefont
  {Zhou}},\ }\href {\doibase 10.1103/PhysRevD.100.095003} {\bibfield  {journal}
  {\bibinfo  {journal} {Phys. Rev.}\ }\textbf {\bibinfo {volume} {D100}},\
  \bibinfo {pages} {095003} (\bibinfo {year} {2019})},\ \Eprint
  {http://arxiv.org/abs/1808.00010} {arXiv:1808.00010 [hep-ph]} \BibitemShut
  {NoStop}%
\bibitem [{\citenamefont {Bi}\ \emph {et~al.}(2019)\citenamefont {Bi},
  \citenamefont {Zhang},\ and\ \citenamefont {Zhou}}]{Bi:2019phv}%
  \BibitemOpen
  \bibfield  {author} {\bibinfo {author} {\bibfnamefont {Q.}~\bibnamefont
  {Bi}}, \bibinfo {author} {\bibfnamefont {C.}~\bibnamefont {Zhang}}, \ and\
  \bibinfo {author} {\bibfnamefont {S.-Y.}\ \bibnamefont {Zhou}},\ }\href
  {\doibase 10.1007/JHEP06(2019)137} {\bibfield  {journal} {\bibinfo  {journal}
  {JHEP}\ }\textbf {\bibinfo {volume} {06}},\ \bibinfo {pages} {137} (\bibinfo
  {year} {2019})},\ \Eprint {http://arxiv.org/abs/1902.08977} {arXiv:1902.08977
  [hep-ph]} \BibitemShut {NoStop}%
\bibitem [{\citenamefont {Bellazzini}\ and\ \citenamefont
  {Riva}(2018)}]{Bellazzini:2018paj}%
  \BibitemOpen
  \bibfield  {author} {\bibinfo {author} {\bibfnamefont {B.}~\bibnamefont
  {Bellazzini}}\ and\ \bibinfo {author} {\bibfnamefont {F.}~\bibnamefont
  {Riva}},\ }\href {\doibase 10.1103/PhysRevD.98.095021} {\bibfield  {journal}
  {\bibinfo  {journal} {Phys. Rev. D}\ }\textbf {\bibinfo {volume} {98}},\
  \bibinfo {pages} {095021} (\bibinfo {year} {2018})},\ \Eprint
  {http://arxiv.org/abs/1806.09640} {arXiv:1806.09640 [hep-ph]} \BibitemShut
  {NoStop}%
\bibitem [{\citenamefont {Remmen}\ and\ \citenamefont
  {Rodd}(2019)}]{Remmen:2019cyz}%
  \BibitemOpen
  \bibfield  {author} {\bibinfo {author} {\bibfnamefont {G.~N.}\ \bibnamefont
  {Remmen}}\ and\ \bibinfo {author} {\bibfnamefont {N.~L.}\ \bibnamefont
  {Rodd}},\ }\href {\doibase 10.1007/JHEP12(2019)032} {\bibfield  {journal}
  {\bibinfo  {journal} {JHEP}\ }\textbf {\bibinfo {volume} {12}},\ \bibinfo
  {pages} {032} (\bibinfo {year} {2019})},\ \Eprint
  {http://arxiv.org/abs/1908.09845} {arXiv:1908.09845 [hep-ph]} \BibitemShut
  {NoStop}%
\bibitem [{\citenamefont {Remmen}\ and\ \citenamefont
  {Rodd}(2020)}]{Remmen:2020vts}%
  \BibitemOpen
  \bibfield  {author} {\bibinfo {author} {\bibfnamefont {G.~N.}\ \bibnamefont
  {Remmen}}\ and\ \bibinfo {author} {\bibfnamefont {N.~L.}\ \bibnamefont
  {Rodd}},\ }\href@noop {} {\  (\bibinfo {year} {2020})},\ \Eprint
  {http://arxiv.org/abs/2004.02885} {arXiv:2004.02885 [hep-ph]} \BibitemShut
  {NoStop}%
\bibitem [{\citenamefont {de~Rham}\ \emph {et~al.}(2019)\citenamefont
  {de~Rham}, \citenamefont {Melville}, \citenamefont {Tolley},\ and\
  \citenamefont {Zhou}}]{deRham:2018qqo}%
  \BibitemOpen
  \bibfield  {author} {\bibinfo {author} {\bibfnamefont {C.}~\bibnamefont
  {de~Rham}}, \bibinfo {author} {\bibfnamefont {S.}~\bibnamefont {Melville}},
  \bibinfo {author} {\bibfnamefont {A.~J.}\ \bibnamefont {Tolley}}, \ and\
  \bibinfo {author} {\bibfnamefont {S.-Y.}\ \bibnamefont {Zhou}},\ }\href
  {\doibase 10.1007/JHEP03(2019)182} {\bibfield  {journal} {\bibinfo  {journal}
  {JHEP}\ }\textbf {\bibinfo {volume} {03}},\ \bibinfo {pages} {182} (\bibinfo
  {year} {2019})},\ \Eprint {http://arxiv.org/abs/1804.10624} {arXiv:1804.10624
  [hep-th]} \BibitemShut {NoStop}%
\bibitem [{\citenamefont {de~Rham}\ \emph
  {et~al.}(2017{\natexlab{b}})\citenamefont {de~Rham}, \citenamefont
  {Melville}, \citenamefont {Tolley},\ and\ \citenamefont
  {Zhou}}]{deRham:2017imi}%
  \BibitemOpen
  \bibfield  {author} {\bibinfo {author} {\bibfnamefont {C.}~\bibnamefont
  {de~Rham}}, \bibinfo {author} {\bibfnamefont {S.}~\bibnamefont {Melville}},
  \bibinfo {author} {\bibfnamefont {A.~J.}\ \bibnamefont {Tolley}}, \ and\
  \bibinfo {author} {\bibfnamefont {S.-Y.}\ \bibnamefont {Zhou}},\ }\href
  {\doibase 10.1007/JHEP09(2017)072} {\bibfield  {journal} {\bibinfo  {journal}
  {JHEP}\ }\textbf {\bibinfo {volume} {09}},\ \bibinfo {pages} {072} (\bibinfo
  {year} {2017}{\natexlab{b}})},\ \Eprint {http://arxiv.org/abs/1702.08577}
  {arXiv:1702.08577 [hep-th]} \BibitemShut {NoStop}%
\bibitem [{\citenamefont {Baumann}\ \emph {et~al.}(2016)\citenamefont
  {Baumann}, \citenamefont {Green}, \citenamefont {Lee},\ and\ \citenamefont
  {Porto}}]{Baumann:2015nta}%
  \BibitemOpen
  \bibfield  {author} {\bibinfo {author} {\bibfnamefont {D.}~\bibnamefont
  {Baumann}}, \bibinfo {author} {\bibfnamefont {D.}~\bibnamefont {Green}},
  \bibinfo {author} {\bibfnamefont {H.}~\bibnamefont {Lee}}, \ and\ \bibinfo
  {author} {\bibfnamefont {R.~A.}\ \bibnamefont {Porto}},\ }\href {\doibase
  10.1103/PhysRevD.93.023523} {\bibfield  {journal} {\bibinfo  {journal} {Phys.
  Rev.}\ }\textbf {\bibinfo {volume} {D93}},\ \bibinfo {pages} {023523}
  (\bibinfo {year} {2016})},\ \Eprint {http://arxiv.org/abs/1502.07304}
  {arXiv:1502.07304 [hep-th]} \BibitemShut {NoStop}%
\bibitem [{\citenamefont {Bellazzini}\ \emph {et~al.}(2016)\citenamefont
  {Bellazzini}, \citenamefont {Cheung},\ and\ \citenamefont
  {Remmen}}]{Bellazzini:2015cra}%
  \BibitemOpen
  \bibfield  {author} {\bibinfo {author} {\bibfnamefont {B.}~\bibnamefont
  {Bellazzini}}, \bibinfo {author} {\bibfnamefont {C.}~\bibnamefont {Cheung}},
  \ and\ \bibinfo {author} {\bibfnamefont {G.~N.}\ \bibnamefont {Remmen}},\
  }\href {\doibase 10.1103/PhysRevD.93.064076} {\bibfield  {journal} {\bibinfo
  {journal} {Phys. Rev.}\ }\textbf {\bibinfo {volume} {D93}},\ \bibinfo {pages}
  {064076} (\bibinfo {year} {2016})},\ \Eprint
  {http://arxiv.org/abs/1509.00851} {arXiv:1509.00851 [hep-th]} \BibitemShut
  {NoStop}%
\bibitem [{\citenamefont {Cheung}\ and\ \citenamefont
  {Remmen}(2016)}]{Cheung:2016yqr}%
  \BibitemOpen
  \bibfield  {author} {\bibinfo {author} {\bibfnamefont {C.}~\bibnamefont
  {Cheung}}\ and\ \bibinfo {author} {\bibfnamefont {G.~N.}\ \bibnamefont
  {Remmen}},\ }\href {\doibase 10.1007/JHEP04(2016)002} {\bibfield  {journal}
  {\bibinfo  {journal} {JHEP}\ }\textbf {\bibinfo {volume} {04}},\ \bibinfo
  {pages} {002} (\bibinfo {year} {2016})},\ \Eprint
  {http://arxiv.org/abs/1601.04068} {arXiv:1601.04068 [hep-th]} \BibitemShut
  {NoStop}%
\bibitem [{\citenamefont {Cheung}\ and\ \citenamefont
  {Remmen}(2017)}]{Cheung:2016wjt}%
  \BibitemOpen
  \bibfield  {author} {\bibinfo {author} {\bibfnamefont {C.}~\bibnamefont
  {Cheung}}\ and\ \bibinfo {author} {\bibfnamefont {G.~N.}\ \bibnamefont
  {Remmen}},\ }\href {\doibase 10.1103/PhysRevLett.118.051601} {\bibfield
  {journal} {\bibinfo  {journal} {Phys. Rev. Lett.}\ }\textbf {\bibinfo
  {volume} {118}},\ \bibinfo {pages} {051601} (\bibinfo {year} {2017})},\
  \Eprint {http://arxiv.org/abs/1608.02942} {arXiv:1608.02942 [hep-th]}
  \BibitemShut {NoStop}%
\bibitem [{\citenamefont {Bellazzini}\ \emph {et~al.}(2018)\citenamefont
  {Bellazzini}, \citenamefont {Riva}, \citenamefont {Serra},\ and\
  \citenamefont {Sgarlata}}]{Bellazzini:2017fep}%
  \BibitemOpen
  \bibfield  {author} {\bibinfo {author} {\bibfnamefont {B.}~\bibnamefont
  {Bellazzini}}, \bibinfo {author} {\bibfnamefont {F.}~\bibnamefont {Riva}},
  \bibinfo {author} {\bibfnamefont {J.}~\bibnamefont {Serra}}, \ and\ \bibinfo
  {author} {\bibfnamefont {F.}~\bibnamefont {Sgarlata}},\ }\href {\doibase
  10.1103/PhysRevLett.120.161101} {\bibfield  {journal} {\bibinfo  {journal}
  {Phys. Rev. Lett.}\ }\textbf {\bibinfo {volume} {120}},\ \bibinfo {pages}
  {161101} (\bibinfo {year} {2018})},\ \Eprint
  {http://arxiv.org/abs/1710.02539} {arXiv:1710.02539 [hep-th]} \BibitemShut
  {NoStop}%
\bibitem [{\citenamefont {Bonifacio}\ \emph {et~al.}(2016)\citenamefont
  {Bonifacio}, \citenamefont {Hinterbichler},\ and\ \citenamefont
  {Rosen}}]{Bonifacio:2016wcb}%
  \BibitemOpen
  \bibfield  {author} {\bibinfo {author} {\bibfnamefont {J.}~\bibnamefont
  {Bonifacio}}, \bibinfo {author} {\bibfnamefont {K.}~\bibnamefont
  {Hinterbichler}}, \ and\ \bibinfo {author} {\bibfnamefont {R.~A.}\
  \bibnamefont {Rosen}},\ }\href {\doibase 10.1103/PhysRevD.94.104001}
  {\bibfield  {journal} {\bibinfo  {journal} {Phys. Rev.}\ }\textbf {\bibinfo
  {volume} {D94}},\ \bibinfo {pages} {104001} (\bibinfo {year} {2016})},\
  \Eprint {http://arxiv.org/abs/1607.06084} {arXiv:1607.06084 [hep-th]}
  \BibitemShut {NoStop}%
\bibitem [{\citenamefont {Hinterbichler}\ \emph {et~al.}(2018)\citenamefont
  {Hinterbichler}, \citenamefont {Joyce},\ and\ \citenamefont
  {Rosen}}]{Hinterbichler:2017qyt}%
  \BibitemOpen
  \bibfield  {author} {\bibinfo {author} {\bibfnamefont {K.}~\bibnamefont
  {Hinterbichler}}, \bibinfo {author} {\bibfnamefont {A.}~\bibnamefont
  {Joyce}}, \ and\ \bibinfo {author} {\bibfnamefont {R.~A.}\ \bibnamefont
  {Rosen}},\ }\href {\doibase 10.1007/JHEP03(2018)051} {\bibfield  {journal}
  {\bibinfo  {journal} {JHEP}\ }\textbf {\bibinfo {volume} {03}},\ \bibinfo
  {pages} {051} (\bibinfo {year} {2018})},\ \Eprint
  {http://arxiv.org/abs/1708.05716} {arXiv:1708.05716 [hep-th]} \BibitemShut
  {NoStop}%
\bibitem [{\citenamefont {Bonifacio}\ \emph {et~al.}(2018)\citenamefont
  {Bonifacio}, \citenamefont {Hinterbichler}, \citenamefont {Joyce},\ and\
  \citenamefont {Rosen}}]{Bonifacio:2017nnt}%
  \BibitemOpen
  \bibfield  {author} {\bibinfo {author} {\bibfnamefont {J.}~\bibnamefont
  {Bonifacio}}, \bibinfo {author} {\bibfnamefont {K.}~\bibnamefont
  {Hinterbichler}}, \bibinfo {author} {\bibfnamefont {A.}~\bibnamefont
  {Joyce}}, \ and\ \bibinfo {author} {\bibfnamefont {R.~A.}\ \bibnamefont
  {Rosen}},\ }\href {\doibase 10.1007/JHEP06(2018)075} {\bibfield  {journal}
  {\bibinfo  {journal} {JHEP}\ }\textbf {\bibinfo {volume} {06}},\ \bibinfo
  {pages} {075} (\bibinfo {year} {2018})},\ \Eprint
  {http://arxiv.org/abs/1712.10020} {arXiv:1712.10020 [hep-th]} \BibitemShut
  {NoStop}%
\bibitem [{\citenamefont {Bellazzini}\ \emph {et~al.}(2017)\citenamefont
  {Bellazzini}, \citenamefont {Riva}, \citenamefont {Serra},\ and\
  \citenamefont {Sgarlata}}]{Bellazzini:2017bkb}%
  \BibitemOpen
  \bibfield  {author} {\bibinfo {author} {\bibfnamefont {B.}~\bibnamefont
  {Bellazzini}}, \bibinfo {author} {\bibfnamefont {F.}~\bibnamefont {Riva}},
  \bibinfo {author} {\bibfnamefont {J.}~\bibnamefont {Serra}}, \ and\ \bibinfo
  {author} {\bibfnamefont {F.}~\bibnamefont {Sgarlata}},\ }\href {\doibase
  10.1007/JHEP11(2017)020} {\bibfield  {journal} {\bibinfo  {journal} {JHEP}\
  }\textbf {\bibinfo {volume} {11}},\ \bibinfo {pages} {020} (\bibinfo {year}
  {2017})},\ \Eprint {http://arxiv.org/abs/1706.03070} {arXiv:1706.03070
  [hep-ph]} \BibitemShut {NoStop}%
\bibitem [{\citenamefont {Bonifacio}\ and\ \citenamefont
  {Hinterbichler}(2018)}]{Bonifacio:2018vzv}%
  \BibitemOpen
  \bibfield  {author} {\bibinfo {author} {\bibfnamefont {J.}~\bibnamefont
  {Bonifacio}}\ and\ \bibinfo {author} {\bibfnamefont {K.}~\bibnamefont
  {Hinterbichler}},\ }\href {\doibase 10.1103/PhysRevD.98.045003} {\bibfield
  {journal} {\bibinfo  {journal} {Phys. Rev.}\ }\textbf {\bibinfo {volume}
  {D98}},\ \bibinfo {pages} {045003} (\bibinfo {year} {2018})},\ \Eprint
  {http://arxiv.org/abs/1804.08686} {arXiv:1804.08686 [hep-th]} \BibitemShut
  {NoStop}%
\bibitem [{\citenamefont {Bellazzini}\ \emph {et~al.}(2019)\citenamefont
  {Bellazzini}, \citenamefont {Lewandowski},\ and\ \citenamefont
  {Serra}}]{Bellazzini:2019xts}%
  \BibitemOpen
  \bibfield  {author} {\bibinfo {author} {\bibfnamefont {B.}~\bibnamefont
  {Bellazzini}}, \bibinfo {author} {\bibfnamefont {M.}~\bibnamefont
  {Lewandowski}}, \ and\ \bibinfo {author} {\bibfnamefont {J.}~\bibnamefont
  {Serra}},\ }\href {\doibase 10.1103/PhysRevLett.123.251103} {\bibfield
  {journal} {\bibinfo  {journal} {Phys. Rev. Lett.}\ }\textbf {\bibinfo
  {volume} {123}},\ \bibinfo {pages} {251103} (\bibinfo {year} {2019})},\
  \Eprint {http://arxiv.org/abs/1902.03250} {arXiv:1902.03250 [hep-th]}
  \BibitemShut {NoStop}%
\bibitem [{\citenamefont {Melville}\ and\ \citenamefont
  {Noller}(2020)}]{Melville:2019wyy}%
  \BibitemOpen
  \bibfield  {author} {\bibinfo {author} {\bibfnamefont {S.}~\bibnamefont
  {Melville}}\ and\ \bibinfo {author} {\bibfnamefont {J.}~\bibnamefont
  {Noller}},\ }\href {\doibase 10.1103/PhysRevD.101.021502} {\bibfield
  {journal} {\bibinfo  {journal} {Phys. Rev.}\ }\textbf {\bibinfo {volume}
  {D101}},\ \bibinfo {pages} {021502} (\bibinfo {year} {2020})},\ \Eprint
  {http://arxiv.org/abs/1904.05874} {arXiv:1904.05874 [astro-ph.CO]}
  \BibitemShut {NoStop}%
\bibitem [{\citenamefont {de~Rham}\ and\ \citenamefont
  {Tolley}(2019)}]{deRham:2019ctd}%
  \BibitemOpen
  \bibfield  {author} {\bibinfo {author} {\bibfnamefont {C.}~\bibnamefont
  {de~Rham}}\ and\ \bibinfo {author} {\bibfnamefont {A.~J.}\ \bibnamefont
  {Tolley}},\ }\href@noop {} {\  (\bibinfo {year} {2019})},\ \Eprint
  {http://arxiv.org/abs/1909.00881} {arXiv:1909.00881 [hep-th]} \BibitemShut
  {NoStop}%
\bibitem [{\citenamefont {Alberte}\ \emph
  {et~al.}(2019{\natexlab{a}})\citenamefont {Alberte}, \citenamefont {de~Rham},
  \citenamefont {Momeni}, \citenamefont {Rumbutis},\ and\ \citenamefont
  {Tolley}}]{Alberte:2019xfh}%
  \BibitemOpen
  \bibfield  {author} {\bibinfo {author} {\bibfnamefont {L.}~\bibnamefont
  {Alberte}}, \bibinfo {author} {\bibfnamefont {C.}~\bibnamefont {de~Rham}},
  \bibinfo {author} {\bibfnamefont {A.}~\bibnamefont {Momeni}}, \bibinfo
  {author} {\bibfnamefont {J.}~\bibnamefont {Rumbutis}}, \ and\ \bibinfo
  {author} {\bibfnamefont {A.~J.}\ \bibnamefont {Tolley}},\ }\href@noop {} {\
  (\bibinfo {year} {2019}{\natexlab{a}})},\ \Eprint
  {http://arxiv.org/abs/1910.11799} {arXiv:1910.11799 [hep-th]} \BibitemShut
  {NoStop}%
\bibitem [{\citenamefont {Alberte}\ \emph
  {et~al.}(2019{\natexlab{b}})\citenamefont {Alberte}, \citenamefont {de~Rham},
  \citenamefont {Momeni}, \citenamefont {Rumbutis},\ and\ \citenamefont
  {Tolley}}]{Alberte:2019zhd}%
  \BibitemOpen
  \bibfield  {author} {\bibinfo {author} {\bibfnamefont {L.}~\bibnamefont
  {Alberte}}, \bibinfo {author} {\bibfnamefont {C.}~\bibnamefont {de~Rham}},
  \bibinfo {author} {\bibfnamefont {A.}~\bibnamefont {Momeni}}, \bibinfo
  {author} {\bibfnamefont {J.}~\bibnamefont {Rumbutis}}, \ and\ \bibinfo
  {author} {\bibfnamefont {A.~J.}\ \bibnamefont {Tolley}},\ }\href@noop {} {\
  (\bibinfo {year} {2019}{\natexlab{b}})},\ \Eprint
  {http://arxiv.org/abs/1912.10018} {arXiv:1912.10018 [hep-th]} \BibitemShut
  {NoStop}%
\bibitem [{\citenamefont {Ye}\ and\ \citenamefont {Piao}(2019)}]{Ye:2019oxx}%
  \BibitemOpen
  \bibfield  {author} {\bibinfo {author} {\bibfnamefont {G.}~\bibnamefont
  {Ye}}\ and\ \bibinfo {author} {\bibfnamefont {Y.-S.}\ \bibnamefont {Piao}},\
  }\href@noop {} {\  (\bibinfo {year} {2019})},\ \Eprint
  {http://arxiv.org/abs/1908.08644} {arXiv:1908.08644 [hep-th]} \BibitemShut
  {NoStop}%
\bibitem [{\citenamefont {Herrero-Valea}\ \emph {et~al.}(2019)\citenamefont
  {Herrero-Valea}, \citenamefont {Timiryasov},\ and\ \citenamefont
  {Tokareva}}]{Herrero-Valea:2019hde}%
  \BibitemOpen
  \bibfield  {author} {\bibinfo {author} {\bibfnamefont {M.}~\bibnamefont
  {Herrero-Valea}}, \bibinfo {author} {\bibfnamefont {I.}~\bibnamefont
  {Timiryasov}}, \ and\ \bibinfo {author} {\bibfnamefont {A.}~\bibnamefont
  {Tokareva}},\ }\href {\doibase 10.1088/1475-7516/2019/11/042} {\  (\bibinfo
  {year} {2019}),\ 10.1088/1475-7516/2019/11/042},\ \Eprint
  {http://arxiv.org/abs/1905.08816} {arXiv:1905.08816 [hep-ph]} \BibitemShut
  {NoStop}%
\bibitem [{\citenamefont {Wang}\ \emph {et~al.}(2020)\citenamefont {Wang},
  \citenamefont {Guo}, \citenamefont {Zhang},\ and\ \citenamefont
  {Zhou}}]{Wang:2020jxr}%
  \BibitemOpen
  \bibfield  {author} {\bibinfo {author} {\bibfnamefont {Y.-J.}\ \bibnamefont
  {Wang}}, \bibinfo {author} {\bibfnamefont {F.-K.}\ \bibnamefont {Guo}},
  \bibinfo {author} {\bibfnamefont {C.}~\bibnamefont {Zhang}}, \ and\ \bibinfo
  {author} {\bibfnamefont {S.-Y.}\ \bibnamefont {Zhou}},\ }\href@noop {} {\
  (\bibinfo {year} {2020})},\ \Eprint {http://arxiv.org/abs/2004.03992}
  {arXiv:2004.03992 [hep-ph]} \BibitemShut {NoStop}%
\bibitem [{\citenamefont {Henning}\ \emph {et~al.}(2017)\citenamefont
  {Henning}, \citenamefont {Lu}, \citenamefont {Melia},\ and\ \citenamefont
  {Murayama}}]{Henning:2015alf}%
  \BibitemOpen
  \bibfield  {author} {\bibinfo {author} {\bibfnamefont {B.}~\bibnamefont
  {Henning}}, \bibinfo {author} {\bibfnamefont {X.}~\bibnamefont {Lu}},
  \bibinfo {author} {\bibfnamefont {T.}~\bibnamefont {Melia}}, \ and\ \bibinfo
  {author} {\bibfnamefont {H.}~\bibnamefont {Murayama}},\ }\href {\doibase
  10.1007/JHEP08(2017)016} {\bibfield  {journal} {\bibinfo  {journal} {JHEP}\
  }\textbf {\bibinfo {volume} {08}},\ \bibinfo {pages} {016} (\bibinfo {year}
  {2017})},\ \bibinfo {note} {[Erratum: JHEP 09, 019 (2019)]},\ \Eprint
  {http://arxiv.org/abs/1512.03433} {arXiv:1512.03433 [hep-ph]} \BibitemShut
  {NoStop}%
\bibitem [{\citenamefont {Murphy}(2020)}]{Murphy:2020rsh}%
  \BibitemOpen
  \bibfield  {author} {\bibinfo {author} {\bibfnamefont {C.~W.}\ \bibnamefont
  {Murphy}},\ }\href@noop {} {\  (\bibinfo {year} {2020})},\ \Eprint
  {http://arxiv.org/abs/2005.00059} {arXiv:2005.00059 [hep-ph]} \BibitemShut
  {NoStop}%
\bibitem [{\citenamefont {Li}\ \emph {et~al.}(2020)\citenamefont {Li},
  \citenamefont {Ren}, \citenamefont {Shu}, \citenamefont {Xiao}, \citenamefont
  {Yu},\ and\ \citenamefont {Zheng}}]{Li:2020gnx}%
  \BibitemOpen
  \bibfield  {author} {\bibinfo {author} {\bibfnamefont {H.-L.}\ \bibnamefont
  {Li}}, \bibinfo {author} {\bibfnamefont {Z.}~\bibnamefont {Ren}}, \bibinfo
  {author} {\bibfnamefont {J.}~\bibnamefont {Shu}}, \bibinfo {author}
  {\bibfnamefont {M.-L.}\ \bibnamefont {Xiao}}, \bibinfo {author}
  {\bibfnamefont {J.-H.}\ \bibnamefont {Yu}}, \ and\ \bibinfo {author}
  {\bibfnamefont {Y.-H.}\ \bibnamefont {Zheng}},\ }\href@noop {} {\  (\bibinfo
  {year} {2020})},\ \Eprint {http://arxiv.org/abs/2005.00008} {arXiv:2005.00008
  [hep-ph]} \BibitemShut {NoStop}%
\bibitem [{\citenamefont {Liu}\ \emph {et~al.}(2016)\citenamefont {Liu},
  \citenamefont {Pomarol}, \citenamefont {Rattazzi},\ and\ \citenamefont
  {Riva}}]{Liu:2016idz}%
  \BibitemOpen
  \bibfield  {author} {\bibinfo {author} {\bibfnamefont {D.}~\bibnamefont
  {Liu}}, \bibinfo {author} {\bibfnamefont {A.}~\bibnamefont {Pomarol}},
  \bibinfo {author} {\bibfnamefont {R.}~\bibnamefont {Rattazzi}}, \ and\
  \bibinfo {author} {\bibfnamefont {F.}~\bibnamefont {Riva}},\ }\href {\doibase
  10.1007/JHEP11(2016)141} {\bibfield  {journal} {\bibinfo  {journal} {JHEP}\
  }\textbf {\bibinfo {volume} {11}},\ \bibinfo {pages} {141} (\bibinfo {year}
  {2016})},\ \Eprint {http://arxiv.org/abs/1603.03064} {arXiv:1603.03064
  [hep-ph]} \BibitemShut {NoStop}%
\bibitem [{\citenamefont {Azatov}\ \emph {et~al.}(2017)\citenamefont {Azatov},
  \citenamefont {Contino}, \citenamefont {Machado},\ and\ \citenamefont
  {Riva}}]{Azatov:2016sqh}%
  \BibitemOpen
  \bibfield  {author} {\bibinfo {author} {\bibfnamefont {A.}~\bibnamefont
  {Azatov}}, \bibinfo {author} {\bibfnamefont {R.}~\bibnamefont {Contino}},
  \bibinfo {author} {\bibfnamefont {C.~S.}\ \bibnamefont {Machado}}, \ and\
  \bibinfo {author} {\bibfnamefont {F.}~\bibnamefont {Riva}},\ }\href {\doibase
  10.1103/PhysRevD.95.065014} {\bibfield  {journal} {\bibinfo  {journal} {Phys.
  Rev. D}\ }\textbf {\bibinfo {volume} {95}},\ \bibinfo {pages} {065014}
  (\bibinfo {year} {2017})},\ \Eprint {http://arxiv.org/abs/1607.05236}
  {arXiv:1607.05236 [hep-ph]} \BibitemShut {NoStop}%
\bibitem [{\citenamefont {Ellis}\ and\ \citenamefont
  {Ge}(2018)}]{Ellis:2018cos}%
  \BibitemOpen
  \bibfield  {author} {\bibinfo {author} {\bibfnamefont {J.}~\bibnamefont
  {Ellis}}\ and\ \bibinfo {author} {\bibfnamefont {S.-F.}\ \bibnamefont {Ge}},\
  }\href {\doibase 10.1103/PhysRevLett.121.041801} {\bibfield  {journal}
  {\bibinfo  {journal} {Phys. Rev. Lett.}\ }\textbf {\bibinfo {volume} {121}},\
  \bibinfo {pages} {041801} (\bibinfo {year} {2018})},\ \Eprint
  {http://arxiv.org/abs/1802.02416} {arXiv:1802.02416 [hep-ph]} \BibitemShut
  {NoStop}%
\bibitem [{\citenamefont {Hays}\ \emph {et~al.}(2019)\citenamefont {Hays},
  \citenamefont {Martin}, \citenamefont {Sanz},\ and\ \citenamefont
  {Setford}}]{Hays:2018zze}%
  \BibitemOpen
  \bibfield  {author} {\bibinfo {author} {\bibfnamefont {C.}~\bibnamefont
  {Hays}}, \bibinfo {author} {\bibfnamefont {A.}~\bibnamefont {Martin}},
  \bibinfo {author} {\bibfnamefont {V.}~\bibnamefont {Sanz}}, \ and\ \bibinfo
  {author} {\bibfnamefont {J.}~\bibnamefont {Setford}},\ }\href {\doibase
  10.1007/JHEP02(2019)123} {\bibfield  {journal} {\bibinfo  {journal} {JHEP}\
  }\textbf {\bibinfo {volume} {02}},\ \bibinfo {pages} {123} (\bibinfo {year}
  {2019})},\ \Eprint {http://arxiv.org/abs/1808.00442} {arXiv:1808.00442
  [hep-ph]} \BibitemShut {NoStop}%
\bibitem [{\citenamefont {Ellis}\ \emph {et~al.}(2020)\citenamefont {Ellis},
  \citenamefont {Ge}, \citenamefont {He},\ and\ \citenamefont
  {Xiao}}]{Ellis:2019zex}%
  \BibitemOpen
  \bibfield  {author} {\bibinfo {author} {\bibfnamefont {J.}~\bibnamefont
  {Ellis}}, \bibinfo {author} {\bibfnamefont {S.-F.}\ \bibnamefont {Ge}},
  \bibinfo {author} {\bibfnamefont {H.-J.}\ \bibnamefont {He}}, \ and\ \bibinfo
  {author} {\bibfnamefont {R.-Q.}\ \bibnamefont {Xiao}},\ }\href {\doibase
  10.1088/1674-1137/44/6/063106} {\bibfield  {journal} {\bibinfo  {journal}
  {Chin. Phys. C}\ }\textbf {\bibinfo {volume} {44}},\ \bibinfo {pages}
  {063106} (\bibinfo {year} {2020})},\ \Eprint
  {http://arxiv.org/abs/1902.06631} {arXiv:1902.06631 [hep-ph]} \BibitemShut
  {NoStop}%
\bibitem [{\citenamefont {Alioli}\ \emph {et~al.}(2020)\citenamefont {Alioli},
  \citenamefont {Boughezal}, \citenamefont {Mereghetti},\ and\ \citenamefont
  {Petriello}}]{Alioli:2020kez}%
  \BibitemOpen
  \bibfield  {author} {\bibinfo {author} {\bibfnamefont {S.}~\bibnamefont
  {Alioli}}, \bibinfo {author} {\bibfnamefont {R.}~\bibnamefont {Boughezal}},
  \bibinfo {author} {\bibfnamefont {E.}~\bibnamefont {Mereghetti}}, \ and\
  \bibinfo {author} {\bibfnamefont {F.}~\bibnamefont {Petriello}},\ }\href@noop
  {} {\  (\bibinfo {year} {2020})},\ \Eprint {http://arxiv.org/abs/2003.11615}
  {arXiv:2003.11615 [hep-ph]} \BibitemShut {NoStop}%
\bibitem [{\citenamefont {Sirunyan}\ \emph {et~al.}(2019)\citenamefont
  {Sirunyan} \emph {et~al.}}]{Sirunyan:2019der}%
  \BibitemOpen
  \bibfield  {author} {\bibinfo {author} {\bibfnamefont {A.~M.}\ \bibnamefont
  {Sirunyan}} \emph {et~al.} (\bibinfo {collaboration} {CMS}),\ }\href
  {\doibase 10.1016/j.physletb.2019.134985} {\bibfield  {journal} {\bibinfo
  {journal} {Phys. Lett. B}\ }\textbf {\bibinfo {volume} {798}},\ \bibinfo
  {pages} {134985} (\bibinfo {year} {2019})},\ \Eprint
  {http://arxiv.org/abs/1905.07445} {arXiv:1905.07445 [hep-ex]} \BibitemShut
  {NoStop}%
\bibitem [{\citenamefont {{CMS Collaboration}}(2020)}]{CMS:2020meo}%
  \BibitemOpen
  \bibfield  {author} {\bibinfo {author} {\bibnamefont {{CMS Collaboration}}}
  (\bibinfo {collaboration} {CMS}),\ }\href@noop {} {\  (\bibinfo {year}
  {2020})}\BibitemShut {NoStop}%
\bibitem [{\citenamefont {Sirunyan}\ \emph {et~al.}(2020)\citenamefont
  {Sirunyan} \emph {et~al.}}]{Sirunyan:2020tlu}%
  \BibitemOpen
  \bibfield  {author} {\bibinfo {author} {\bibfnamefont {A.~M.}\ \bibnamefont
  {Sirunyan}} \emph {et~al.} (\bibinfo {collaboration} {CMS}),\ }\href@noop {}
  {\  (\bibinfo {year} {2020})},\ \Eprint {http://arxiv.org/abs/2002.09902}
  {arXiv:2002.09902 [hep-ex]} \BibitemShut {NoStop}%
\bibitem [{\citenamefont {Azzi}\ \emph {et~al.}(2019)\citenamefont {Azzi} \emph
  {et~al.}}]{Azzi:2019yne}%
  \BibitemOpen
  \bibfield  {author} {\bibinfo {author} {\bibfnamefont {P.}~\bibnamefont
  {Azzi}} \emph {et~al.},\ }\enquote {\bibinfo {title} {{Report from Working
  Group 1}: {Standard Model Physics at the HL-LHC and HE-LHC}},}\ in\ \href
  {\doibase 10.23731/CYRM-2019-007.1} {\emph {\bibinfo {booktitle} {{Report on
  the Physics at the HL-LHC,and Perspectives for the HE-LHC}}}},\ Vol.~\bibinfo
  {volume} {7},\ \bibinfo {editor} {edited by\ \bibinfo {editor} {\bibfnamefont
  {A.}~\bibnamefont {Dainese}}, \bibinfo {editor} {\bibfnamefont
  {M.}~\bibnamefont {Mangano}}, \bibinfo {editor} {\bibfnamefont {A.~B.}\
  \bibnamefont {Meyer}}, \bibinfo {editor} {\bibfnamefont {A.}~\bibnamefont
  {Nisati}}, \bibinfo {editor} {\bibfnamefont {G.}~\bibnamefont {Salam}}, \
  and\ \bibinfo {editor} {\bibfnamefont {M.~A.}\ \bibnamefont {Vesterinen}}}\
  (\bibinfo {year} {2019})\ pp.\ \bibinfo {pages} {1--220},\ \Eprint
  {http://arxiv.org/abs/1902.04070} {arXiv:1902.04070 [hep-ph]} \BibitemShut
  {NoStop}%
\bibitem [{\citenamefont {Gupta}(2020)}]{Gupta}%
  \BibitemOpen
  \bibfield  {author} {\bibinfo {author} {\bibfnamefont {R.~S.}\ \bibnamefont
  {Gupta}},\ }\href@noop {} {\enquote {\bibinfo {title} {{Liberating Higgs/EW
  observables at dimension 8}},}\ } (\bibinfo {year} {2020}),\ \bibinfo {note}
  {talk presented at the workshop Higgs and Effective Field Theory 2020.
  \url{https://indico.cern.ch/event/855352/contributions/3759834/attachments/2020293/3377866/heft2020.pdf}}\BibitemShut
  {NoStop}%
\bibitem [{\citenamefont {Krein}\ and\ \citenamefont {Milman}(1940)}]{KM}%
  \BibitemOpen
  \bibfield  {author} {\bibinfo {author} {\bibfnamefont {M.}~\bibnamefont
  {Krein}}\ and\ \bibinfo {author} {\bibfnamefont {D.}~\bibnamefont {Milman}},\
  }\href@noop {} {\bibfield  {journal} {\bibinfo  {journal} {Studia
  Mathematica}\ }\textbf {\bibinfo {volume} {9}},\ \bibinfo {pages} {133}
  (\bibinfo {year} {1940})}\BibitemShut {NoStop}%
\bibitem [{\citenamefont {Andriolo}\ \emph {et~al.}(2020)\citenamefont
  {Andriolo}, \citenamefont {Huang}, \citenamefont {Noumi}, \citenamefont
  {Ooguri},\ and\ \citenamefont {Shiu}}]{Andriolo:2020lul}%
  \BibitemOpen
  \bibfield  {author} {\bibinfo {author} {\bibfnamefont {S.}~\bibnamefont
  {Andriolo}}, \bibinfo {author} {\bibfnamefont {T.-C.}\ \bibnamefont {Huang}},
  \bibinfo {author} {\bibfnamefont {T.}~\bibnamefont {Noumi}}, \bibinfo
  {author} {\bibfnamefont {H.}~\bibnamefont {Ooguri}}, \ and\ \bibinfo {author}
  {\bibfnamefont {G.}~\bibnamefont {Shiu}},\ }\href@noop {} {\  (\bibinfo
  {year} {2020})},\ \Eprint {http://arxiv.org/abs/2004.13721} {arXiv:2004.13721
  [hep-th]} \BibitemShut {NoStop}%
\bibitem [{\citenamefont {Yamashita}\ \emph {et~al.}(2020)\citenamefont
  {Yamashita}, \citenamefont {Zhang},\ and\ \citenamefont {Zhou}}]{longpaper}%
  \BibitemOpen
  \bibfield  {author} {\bibinfo {author} {\bibfnamefont {K.}~\bibnamefont
  {Yamashita}}, \bibinfo {author} {\bibfnamefont {C.}~\bibnamefont {Zhang}}, \
  and\ \bibinfo {author} {\bibfnamefont {S.-Y.}\ \bibnamefont {Zhou}},\
  }\href@noop {} {\  (\bibinfo {year} {2020})},\ \Eprint
  {http://arxiv.org/abs/2009.04490} {arXiv:2009.04490 [hep-ph]} \BibitemShut
  {NoStop}%
\bibitem [{\citenamefont {Minkowski}(1968)}]{Minkowski}%
  \BibitemOpen
  \bibfield  {author} {\bibinfo {author} {\bibfnamefont {H.}~\bibnamefont
  {Minkowski}},\ }\enquote {\bibinfo {title} {Geometry of numbers. (geometrie
  der zahlen.)},}\ \ (\bibinfo  {publisher} {Bibliotheca Mathematica
  Teubneriana. 40. New York, NY: Johnson Reprint Corp. vii, 256 p.},\ \bibinfo
  {year} {1968})\BibitemShut {NoStop}%
\bibitem [{\citenamefont {Weyl}(1935)}]{Weyl}%
  \BibitemOpen
  \bibfield  {author} {\bibinfo {author} {\bibfnamefont {H.}~\bibnamefont
  {Weyl}},\ }\href@noop {} {\bibfield  {journal} {\bibinfo  {journal}
  {Commentarii math. Helvetici}\ }\textbf {\bibinfo {volume} {7}},\ \bibinfo
  {pages} {290} (\bibinfo {year} {1935})}\BibitemShut {NoStop}%
\bibitem [{\citenamefont {Avis}\ and\ \citenamefont {Fukuda}(1992)}]{Avis}%
  \BibitemOpen
  \bibfield  {author} {\bibinfo {author} {\bibfnamefont {D.}~\bibnamefont
  {Avis}}\ and\ \bibinfo {author} {\bibfnamefont {K.}~\bibnamefont {Fukuda}},\
  }\href@noop {} {\bibfield  {journal} {\bibinfo  {journal} {Discrete and
  Computational Geometry}\ }\textbf {\bibinfo {volume} {8}},\ \bibinfo {pages}
  {295} (\bibinfo {year} {1992})}\BibitemShut {NoStop}%
\bibitem [{\citenamefont {Avis}()}]{lrs}%
  \BibitemOpen
  \bibfield  {author} {\bibinfo {author} {\bibfnamefont {D.}~\bibnamefont
  {Avis}},\ }\href@noop {} {}\bibinfo {howpublished}
  {\url{http://cgm.cs.mcgill.ca/~avis/C/lrs.html}}\BibitemShut {NoStop}%
\bibitem [{\citenamefont {Helset}\ \emph {et~al.}(2018)\citenamefont {Helset},
  \citenamefont {Paraskevas},\ and\ \citenamefont {Trott}}]{Helset:2018fgq}%
  \BibitemOpen
  \bibfield  {author} {\bibinfo {author} {\bibfnamefont {A.}~\bibnamefont
  {Helset}}, \bibinfo {author} {\bibfnamefont {M.}~\bibnamefont {Paraskevas}},
  \ and\ \bibinfo {author} {\bibfnamefont {M.}~\bibnamefont {Trott}},\ }\href
  {\doibase 10.1103/PhysRevLett.120.251801} {\bibfield  {journal} {\bibinfo
  {journal} {Phys. Rev. Lett.}\ }\textbf {\bibinfo {volume} {120}},\ \bibinfo
  {pages} {251801} (\bibinfo {year} {2018})},\ \Eprint
  {http://arxiv.org/abs/1803.08001} {arXiv:1803.08001 [hep-ph]} \BibitemShut
  {NoStop}%
\bibitem [{\citenamefont {Low}\ \emph {et~al.}(2010)\citenamefont {Low},
  \citenamefont {Rattazzi},\ and\ \citenamefont {Vichi}}]{Low:2009di}%
  \BibitemOpen
  \bibfield  {author} {\bibinfo {author} {\bibfnamefont {I.}~\bibnamefont
  {Low}}, \bibinfo {author} {\bibfnamefont {R.}~\bibnamefont {Rattazzi}}, \
  and\ \bibinfo {author} {\bibfnamefont {A.}~\bibnamefont {Vichi}},\ }\href
  {\doibase 10.1007/JHEP04(2010)126} {\bibfield  {journal} {\bibinfo  {journal}
  {JHEP}\ }\textbf {\bibinfo {volume} {04}},\ \bibinfo {pages} {126} (\bibinfo
  {year} {2010})},\ \Eprint {http://arxiv.org/abs/0907.5413} {arXiv:0907.5413
  [hep-ph]} \BibitemShut {NoStop}%
\bibitem [{\citenamefont {Degrande}\ \emph {et~al.}(2013)\citenamefont
  {Degrande}, \citenamefont {Eboli}, \citenamefont {Feigl}, \citenamefont
  {Jäger}, \citenamefont {Kilian}, \citenamefont {Mattelaer}, \citenamefont
  {Rauch}, \citenamefont {Reuter}, \citenamefont {Sekulla},\ and\ \citenamefont
  {Wackeroth}}]{Degrande:2013rea}%
  \BibitemOpen
  \bibfield  {author} {\bibinfo {author} {\bibfnamefont {C.}~\bibnamefont
  {Degrande}}, \bibinfo {author} {\bibfnamefont {O.}~\bibnamefont {Eboli}},
  \bibinfo {author} {\bibfnamefont {B.}~\bibnamefont {Feigl}}, \bibinfo
  {author} {\bibfnamefont {B.}~\bibnamefont {Jäger}}, \bibinfo {author}
  {\bibfnamefont {W.}~\bibnamefont {Kilian}}, \bibinfo {author} {\bibfnamefont
  {O.}~\bibnamefont {Mattelaer}}, \bibinfo {author} {\bibfnamefont
  {M.}~\bibnamefont {Rauch}}, \bibinfo {author} {\bibfnamefont
  {J.}~\bibnamefont {Reuter}}, \bibinfo {author} {\bibfnamefont
  {M.}~\bibnamefont {Sekulla}}, \ and\ \bibinfo {author} {\bibfnamefont
  {D.}~\bibnamefont {Wackeroth}},\ }in\ \href
  {http://inspirehep.net/record/1256129/files/arXiv:1309.7890.pdf} {\emph
  {\bibinfo {booktitle} {{Proceedings, 2013 Community Summer Study on the
  Future of U.S. Particle Physics: Snowmass on the Mississippi (CSS2013):
  Minneapolis, MN, USA, July 29-August 6, 2013}}}}\ (\bibinfo {year} {2013})\
  \Eprint {http://arxiv.org/abs/1309.7890} {arXiv:1309.7890 [hep-ph]}
  \BibitemShut {NoStop}%
\bibitem [{Note2()}]{Note2}%
  \BibitemOpen
  \bibinfo {note} {$C_{T,10}$ denotes the coefficient of $O_2^{W^4}$ of \cite
  {Remmen:2019cyz}, multiplied by $g_2^4/4$.}\BibitemShut {Stop}%
\bibitem [{Note3()}]{Note3}%
  \BibitemOpen
  \bibinfo {note} {The EFThedron of \cite {nimahuanghuang} also connects the
  the UV states in a geometric point of view. Convex objects such as cyclic
  polytopes are found to constrain sequences of operators with increasing
  dimensions. Here we consider EFTs endowed with symmetries and focus on
  operators with lowest dimensions.}\BibitemShut {Stop}%
\end{thebibliography}%
\bibliographystyle{apsrev4-1.bst}

\appendix

\setcounter{equation}{14}
\section{APPENDIX}

\sec{The dispersion relation}
Here we present more details about how to get the dispersion relation (2).
Let $\tilde M_{ij\to kl}(s)$ be the forward amplitude $M_{ij\to kl}(s,t=0)$ but
with the pole contributions subtracted out. Using unitarity and analyticity of
the amplitude and Cauchy's integral formula, we can derive a dispersion
relation (see e.g.~[23, 27])
\begin{align}
\label{eq0}
	\tilde M^{ijkl} &\equiv \frac{1}{2}\frac{\ud^2}{\ud s^2}\tilde M_{ij\to kl}(s={M^2}/{2}) 
	+c.c. 
	\\
	&= \int^\infty_{M_{\rm th}^2} \frac{\ud \mu}{2i\pi}\frac{{\rm Disc}M_{ij\to kl}(\mu)}{(\mu-\frac{M^2}{2})^3}  +(j\!\leftrightarrow\! l)+c.c. ,
\end{align}
where the discontinuity of a complex function is defined as ${\rm Disc} A(s)= A(s+i \varepsilon)-A(s-i\varepsilon)$, $M_{\rm th}$ is the threshold scale of the process $ij \to kl$ and $M^2$ is the sum of the four squared masses. Now, for a valid EFT, since we can compute the amplitude in the IR to a desired accuracy within the EFT, we can subtract out the low energy parts of the dispersive integrals
\begin{align}
M^{ijkl}  &\equiv \tilde M^{ijkl} - \int^{(\epsilon\Lambda)^2}_{M_{\rm th}^2} \frac{\ud \mu}{2i\pi}\frac{{\rm Disc}M_{ij\to kl}(\mu)}{(\mu-\frac{M^2}{2})^3} 
	\nonumber \\
	&~~~~~~~~~ -  \int^{(\epsilon\Lambda)^2}_{M_{\rm th}^2} \frac{\ud \mu}{2i\pi}\frac{{\rm Disc}M_{il\to kj}(\mu)}{(\mu-\frac{M^2}{2})^3} +c.c.
	\\
	& = \int^\infty_{(\epsilon\Lambda)^2} \! \frac{\ud \mu}{2i\pi}\frac{{\rm Disc}M_{ij\to kl}(\mu)}{(\mu-\frac{M^2}{2})^3}  +(j\!\leftrightarrow\! l)+c.c.  ,
\end{align}
where $\epsilon\Lambda$ is a scale smaller than $\Lambda$ (for tree-level UV-completions, this scale can be pushed all the way up to the first state lies outside the EFT), but still much larger
than $M_{\rm th}$, so that the denominator of the integrand is
positive.  Using Hermitian analyticity $M_{kl\to ij}^*(s+i\varepsilon) =
M_{ij\to kl}(s-i\varepsilon)$ and the generalized optical theorem $M_{ij\to kl}
- M_{kl\to ij}^*= i\sum'_X M_{ij\to X} M_{kl\to X}^*$, we can then get
Eq.~(2). 

For scatterings with different masses, the forward limit (scattering angle
$\theta=0$) in general does not correspond to $t=0$ where crossing is more
complex and kinematic singularities may incur (see [24] and reference therein).
Also, the total mass squared $M^2$ picks up some $ijkl$ dependence. However,
thanks to the $\epsilon \Lambda$ subtraction, $\mu$ in the dispersion integral
is much greater than the particle masses, our formalism still approximately
applies. When $\mu\gg M^2$, $t=0$ becomes $\theta=0$,
with corrections suppressed by ${\cal O}(M/(\epsilon\Lambda))$. In Eq.~(4),
this would imply that the ERs are the tensors
$P_r^{i(j|k|l)}/[s-M(i,j,k,l)^2/2]^3$ for $s\ge (\epsilon\Lambda)^2$. This
simply smears our original ERs, $P_r^{i(j|k|l)}$, by an amount of at most
$M^2/(\epsilon\Lambda)^2$. These are higher-order effects in an EFT expansion.
In the explicit examples considered in this work, the mass differences are
always negligible.

We also want to mention that in this paper we have focused on the $s^2$ subspace, as this is most accessible experimentally for SMEFT. However, our analysis will be similar for the $s^{2n}$ EFT subspaces $(n=2,3,...)$, which corresponds to taking $s^{2n}$ derivatives in Eq.~(\ref{eq0}).

\sec{Proof of more than elastic positivity}
Here we present more details about how to get the bounds in Eqs.~(10) and
(11) from $T_{1,2}$ given in Eqs.~(12) and
(13), and why these bounds can not be derived from the positivity bounds of scattering between superposed states.  By construction, the $j,l$ indices of $T^{ijkl}_{1,2}$ are symmetrized.
Viewing $ij$ (and $kl$) as one index, $T^{ijkl}_{1,2}$ are both PSD as they have the same positive eigenvalues:
\begin{flalign}
	15, 10, 10, 10, 6, 6, 6, 6, 6, 6, 6, 6, 6, 6, 5, 2, 2, 2, 2, 2,
\nonumber
\end{flalign}
plus 16 zero eigenvalues. Therefore, $T_{1,2}^{ijkl}m^{ij}m^{kl} = T_{1,2}^{ijkl}m^{il}m^{kj}\ge0$.
It follows that $T^{ijkl}_{1,2}M^{ijkl}\ge0$,
which then leads to Eqs.~(10) and (11).

Now we show that $T_{1,2}^{ijkl}\notin \mc Q$, i.e.~the same bounds cannot be
derived from the positivity bounds of the form $u^iv^ju^kv^lM^{ijikl}\ge 0$. To that end,  we need to
show that $T_{1,2}^{ijkl}$ cannot be written as $\sum_a \alpha_a u^i_a v^j_a
u^k_a v^l_a$ with $\alpha_a>0$.   Suppose this can be done for $T_{1}^{ijkl}$. Notice that $T_{1}^{ijkl}
E_b^{ijkl}\!\!=\!\!\sum_a \alpha_a u_a^iv_a^ju_a^kv_a^lE_b^{ijkl}\!\!=\!\!0$
for $b\!\!=\!\!(1,2),(1,3),(3,1),(3,2)$.  Since $E_b^{ijkl}$ are projection operators,
$u_a^iv_a^ju_a^kv_a^lE_b^{ijkl}$ are sums of squares, so for these $b$ values
we have $u_a^iv_a^ju_a^kv_a^lE_b^{ijkl}=0$ for each $a$.  Then $T_{1}^{ijkl}
E_b^{ijkl}\!=\!0$ reduces to a system of quadratic equations for $u,v$, and one can
check explicitly that it has no non-zero solution. Similarly, we can prove that
$T_2^{ijkl}$ can not be written as $\sum_a \alpha_a u^i_a v^j_a
u^k_a v^l_a$ with $\alpha_a>0$, using $E_b^{ijkl}$ for $b=(2,2),(2,3),(3,1),(3,2)$.
So Eqs.~(10) and (11) cannot be derived from
positivity bounds of scattering between $u^i\ket{i}$ and $v^i\ket{j}$.

\end{document}